\documentclass[a4paper,11pt]{article}
\pdfoutput=1 
\usepackage{jcappub} 

\usepackage[T1]{fontenc} 
\usepackage{graphicx}
\usepackage{caption}
\usepackage{subcaption}
\usepackage[numbers,sort&compress]{natbib}
\usepackage[export]{adjustbox}
\usepackage{array}
\usepackage{booktabs}
\usepackage{makecell}
\usepackage[table,dvipsnames]{xcolor}
\usepackage{soul}
\usepackage{upgreek}
\usepackage{wasysym}

\newcommand{\dthetalens}{{\Delta\,\theta_{a,{\rm lens}}^{(n)}}}
\newcommand{\dthetalensS}{{\Delta\,\theta_{a,{\rm lens}}}} 
\newcommand{\gag}{g_{a\gamma}}

\title{Constraining Axion-like Particles through Multi-epoch Monitoring of Strong Gravitational Lenses}

\author[a,1]{Shivani Deshmukh,\note{Corresponding authors.}}
\author[b,c,d,1]{Aritra Basu,}
\author[a,1]{Dominik J. Schwarz,}
\author[e,f]{Yuko Urakawa,}
\author[d]{Sui Ann Mao}

\affiliation[a]{Fakult\"at f\"ur Physik, Universität Bielefeld, Universitätsstraße 25, 33615 Bielefeld, Germany}
\affiliation[b]{Deutsches Zentrum f\"ur Astrophysik, Postplatz 1, 02826 G\"orlitz, Germany}
\affiliation[c]{Th\"uringer Landessternwarte, Sternwarte 5, 07778 Tautenburg, Germany}
\affiliation[d]{Max-Planck-Institut f\"ur Radioastronomie, Auf dem H\"ugel 69, 53121 Bonn, Germany}
\affiliation[e]{Institute of Particle and Nuclear Studies (IPNS), High Energy Accelerator Research Organization
(KEK), Oho 1-1, Tsukuba 305-0801, Japan}
\affiliation[f]{The Graduate University for Advanced Studies (SOKENDAI), Tsukuba 305-0801, Japan}

\emailAdd{sdeshmukh@physik.uni-bielefeld.de}
\emailAdd{aritra.basu@dzastro.de}
\emailAdd{dschwarz@physik.uni-bielefeld.de}

\abstract{
We present new constraints on ultralight axion-like particles (ALPs) through multi-epoch measurements of differential birefringence induced due to a coupling ($\gag$) between the ALP and electromagnetic fields.
Broadband polarimetric observations in the 2--8\,GHz range of the gravitationally lensed system CLASS\,B1152+199 were carried out over five epochs spanning three months with a cadence of roughly 20\,days, and the differential birefringence angle ($\dthetalensS$) between the lensed images were estimated. We also combined an archival observation that effectively increases the span to 9.5\,yr to probe the effect of an oscillating ALP field imprinted as oscillating $\dthetalensS$ over time. Here we present a new technique for combining multi-epoch measurements of $\dthetalensS$ by considering the coherence of the ALP field, such that, $\dthetalensS$ over these observations are related. The time scale of coherence depends on the mass of the ALP field ($m_a$). 
Our results are consistent with non-detection and we constrain $\gag \leq 7.8\times 10^{-12} \,\left( {\rho_{a,\text{em}}}/{20 \text{ GeV cm}^{-3}} \right)^{-1/2}\;\mathrm{GeV}^{-1}$ to $\leq 3.2\times 10^{-8} \,\left( {\rho_{a,\text{em}}}/{20 \text{ GeV cm}^{-3}} \right)^{-1/2}\;\mathrm{GeV}^{-1}$ at 95\% confidence for $m_a$ between $1.6\times 10^{-22}\;\mathrm{eV}$ and $3.8\times 10^{-18}\;\mathrm{eV}$, where $\rho_{a,{\rm em}}$ is the density of the ALP field at emission. This improves over the constraint provided by the CERN Axion Solar Telescope by up to an order of magnitude
in the $m_a$ range $1.6\times 10^{-22}\;\mathrm{eV}$ to $3\times 10^{-21}$\,eV. 
}

\begin{document}

\maketitle
\flushbottom

\section{Introduction} \label{sec:Introduction}

The nature 
of dark matter remains a mystery, even 
close to a century after the first hint for its existence, and more than four decades after its establishment as one of the essential building blocks of today's standard model of cosmology.
First hypothesised based on the high velocity dispersion of galaxies in the Coma cluster \cite{Zwicky1933},
later followed by the observation of
the flatness of galaxy rotation curves
\cite{Rubin1980}, the mass reconstruction of 
galaxy clusters using weak lensing \cite{kaiser1993, Clowe2004}, and the large-scale structure of the Universe \cite{Blumenthal1984, Davis1985} --- all point towards its existence. 
Furthermore, the observed deuterium to hydrogen ratio \cite{Pettini2012} and fluctuations in the cosmic microwave background (CMB) radiation suggest that 
baryonic matter makes up 
at most $5\%$ of the Universe's mass \cite{Planck2020}, establishing strong evidence for the `missing' mass.
Many theories and models of dark matter have been proposed to explain these phenomena. Strong evidence hints towards the existence of elementary particles beyond those of the Standard Model of particle physics, including a plethora of candidates for dark matter  \cite{Bertone2018}. 
The QCD axion, motivated by the \textit{strong CP problem} of the strong interaction \cite{PecceiQuinn1977, PecceiQuinn21977, Wilczek1978, Peccei2008}, and its generalisation, an axion-like particle (ALP)
\cite{Kim:1979if,shifm80,Dine:1981rt,Zhitnitsky:1980, Anselm1982, Preskill1983, Abbott1983, Dine1983, Joerg2010, Ringwald2012, Ringwald2014, Marsh2016}, 
are promising candidates. 
ALPs constitute a general class of pseudo-scalar bosons that could span a broad range of masses and coupling constants. 

Massive efforts are being undertaken to 
detect signatures of axion and ALPs. Interconversion of axions and photons in the presence of magnetic fields, 
the Primakoff effect, has dominated the search efforts \cite{Sisk-Reynes2022, Reynolds2020, Marsh2017, Berg2017, Ning&Safdi2024, Dessert2020}.
Recently, a surge of astrophysical polarisation observations 
to search for ALPs has emerged 
\cite{Fujita2019, Fedderke2019, Chen2020, Chigusa2020, Basu2021, SPT-3G2022, Adachi2024, Adkins2025, Porayko2025}.
These searches exploit the property of ALPs affecting the polarisation of light 
\cite{Maiani1986,Raffelt&Stodolsky1988} emitted from astrophysical sources \cite{Harari&Sikivie1992}. 
Axionic fields give rise to a helicity-dependent dispersion relation for electromagnetic radiation, which leads to a rotation of the plane of linearly polarised light, the 
phenomenon of birefringence.
As the interaction of a time oscillating ALP field with linearly polarised emission results in 
an oscillating rotation of the plane of polarisation, 
monitoring of 
polarised astrophysical sources might allow us to reveal such an oscillation of polarisation angles, provided ALPs exist.

In this work, we focus on the search for ultralight ALPs
and present new, competitive upper limits on their coupling to light, based on new multi-epoch observations of the gravitational lens system CLASS\,B1152+199. This strongly lensed system allows us to observe two polarised images of a quasar across a range of radio frequencies, which we use to robustly constrain the birefringence signatures of ALPs via a novel technique 
first described and applied on the same system \cite{Basu2021}. We extend this technique to multi-epoch monitoring observations, obtained 
with the 
Karl G. Jansky Very Large Array (VLA),
and improve the corresponding constraints on ultralight ALPs, also demonstrating the discovery potential of the method. 
Our analysis probes the ALP density within the region of polarised radio emission associated with the quasar and does not assume that ALPs make up all of the dark matter.

The paper is structured as follows. We start by describing the ALP search technique and its extension 
to combine multiple epochs of observations in section~\ref{sec:ALPtech}. 
In section~\ref{sec:Obs}, we present our observations, outline the calibration and imaging procedure, 
and describe the method for obtaining source polarisation. In section~\ref{sec:constrain} we compute the birefringence angle, and in section~\ref{sec:constrain_ALP} we present the constraints on ALPs obtained in this work. We discuss the implications of monitoring strong gravitational lenses on further constraining ALPs using future facilities such as the LOFAR2.0 and SKA-Mid in section~\ref{sec:discussion}, and conclude in section~\ref{sec:conclusion}. We provide supplementary material in four appendices. Unless otherwise stated, we work in the Heaviside-Lorentz system and use natural units with $c=\hbar=1$.

\section{ALP induced birefringence in strong gravitational lens systems} \label{sec:ALPtech}

In this study, we focus on the search for ALPs in the ultralight mass regime. 
We build upon the novel approach put forth by Basu et al.~\cite{Basu2021} of using differential measurements of polarisation angle from gravitationally lensed, multiple images of a linearly polarised source, such as a quasar, to probe ALPs.
A linearly polarised background quasar undergoes \textit{achromatic birefringence} in the presence of an ALP field \cite{Schwarz2021, Blas2020, Fedderke2019} 
resulting in the rotation of the polarisation angle of the electric vector by an angle $\theta_a$ that depends on the ALP--photon coupling ($\gag$) and the difference of ALP field strength between emission and observations.
Basu et al.~\cite{Basu2021} pointed out that, when that quasar is strongly lensed into multiple images,
the difference of the polarisation angle of the lensed images is a robust, clean probe of \emph{differential birefringence} induced by the ALP field, that is less vulnerable to 
standard observational systematics and astrophysical assumptions than many other methods. Because of the different emission 
time of each of the lensed images, originating from their different 
paths around the lensing object, the birefringence angles for them are expected to be different.  
Schwarz et al.~\cite{Schwarz2021} showed that the birefringence angle remains unaffected for light propagating in the curved space-time of the lensing object in the eikonal approximation (there is a tiny rotation of the plane of polarisation proportional to the deflection angle, which is insignificant compared to the typical measurement uncertainties of polarisation angles).

On solving the Klein-Gordon equation of motion within gravitationally bound structure of dark matter halo, the ALP field oscillates in time (and space) and the period of oscillation ($T_a$) is inversely proportional to ALP mass ($m_a$), $T_a = 2\pi/m_a$.
As a consequence, the birefringence angle should also oscillate.
In the following, we investigate a simple gravitational lens system with two spatially unresolved images.
The rays of light coming from the two images of such a gravitationally lensed system travel different paths from the quasar to an observer which
introduces a time delay between the two images. The observation of lensed images is simultaneous, 
ensuring that the ALP field
at observation is the same for both images. In the presence of an oscillating ALP field at emission, the two images incur different 
birefringence. Therefore,
the difference between the polarisation angles of the images captures the ALP field oscillation at emission, and the ALP field oscillation at the observer is cancelled \cite{Basu2021}. Thus, all relevant time intervals 
are taken in the rest frame of the emission. We transform time intervals from the observer's frame ($\Delta\,t_{\rm o}$) to the rest frame of emission ($\Delta\,t_{\rm em}$) using the relation 
$\Delta\,t_\text{em} = \Delta\,t_\text{o}/(1+z_\text{qso})$,
where $z_\text{qso}$ is the redshift of the lensed quasar for our case.

\subsection{Differential birefringence as a probe of ALP field} \label{Sec:SingleEpoch}

For two strongly lensed images, labeled A and B, the differential polarisation angle is defined as,
\begin{equation}
    \Delta \theta_{a,\text{lens}} = \theta_{\rm 0, A} - \theta_{\rm 0, B} \;.
    \label{eq:difftheta}
\end{equation}
Here, $\theta_{\rm 0, A}$ and $\theta_{\rm 0, B}$ are the intrinsic polarisation angles, corrected for frequency dependent effects, e.g., Faraday rotation, of the lensed images A and B, respectively. 
$\dthetalensS$ obtained at a single epoch is given as \cite{Basu2021},
\begin{equation} \label{Eq:DiffBir}
    \Delta \theta_{a, \text{lens}} = K \sin \left[ \frac{m_a\, \Delta t_\text{d}}{2} \right] \sin \left[ m_at_\text{em} + \tilde{\delta}_\text{em} - \frac{\pi}{2} \right],
\end{equation}
where, the amplitude,
\begin{equation} \label{Eq:K}
    K = 10.04\,{\rm degree}\, \left[ \frac{\rho_{a,\text{em}}}{20 \text{ GeV cm}^{-3}} \right] ^{1/2} \left[ \frac{g_{a\gamma}}{10^{-12} \text{ GeV}^{-1}} \right] \left[ \frac{m_a}{10^{-22} \text{ eV}} \right] ^{-1}.
\end{equation}
Above, $\Delta t_\text{d} = |t_A-t_B|$ denotes the lensing time delay in the emission 
frame, $t_\text{em} = (t_A+t_B)/2$ is the mean time of emission 
of both images, and $\tilde{\delta}_\text{em}$ 
is the phase of the ALP field within a coherent 
patch at the emitting region at that moment. 
$\rho_{a,{\rm em}}$ is the energy density of the ALP field in the emitting region.
Notice that the magnitude of 
$\Delta\,\theta_{a,{\rm lens}}$ is determined by $\Delta t_\text{d}$, and the second sine term indicates that $\Delta \theta_{a,\text{lens}}$ also oscillates with the same period as that of the ALP field, but $- 90$\,degrees 
out of phase with respect to it.
Thus, monitoring of strongly lensed systems to infer the time variation of $\Delta\,\theta_{a,{\rm lens}}$ holds crucial information on an ALP field at emission if it exists.
We would like to emphasise that, the expected differential birefringence signal does not necessarily require ALPs to compose 100\% of dark matter, and therefore detection of an oscillatory signal would address the general question of whether they exist.
Our reference value of an ALP energy density of $20$ GeV cm$^{-3}$ at the core of the quasar, corresponds to a scenario with ALPs making up half of the dark matter, see \cite{Basu2021}.

\subsection{Differential birefringence over multiple epochs} \label{Sec:MultiEpoch}

We further expand on the above technique to incorporate multiple epochs of observations,
and start with generalising eq.~(\ref{Eq:DiffBir}) for the  $n^\text{th}$ epoch as,
\begin{equation} \label{Eq:DiffBir_multi_incoherent}
    \Delta \theta_{a,\text{lens}}^{\,(n)} = K \sin \left[ \frac{m_a\, \Delta t_\text{d}}{2} \right] \sin \left[ m_a\, t_\text{em}^{(n)}  + \delta_\text{em}^{(n)} - \frac{\pi}{2} \right],
\end{equation}
where, $t_\text{em}^{\,(n)}$ and $\delta_{\rm em}^{(n)}$ are the emission time and phase at the $n^\text{th}$ epoch of observation, with time measured in the emission 
frame. Note that, we 
implicitly assume that the amplitude $K$, which depends on 
$\rho_{a,{\rm em}}$, and thus the dark matter density in the core region of the quasar, does not change in any significant way between different epoch of observations. 
For jointly analysing $\dthetalens$, 
we need to consider two scenarios, namely coherent and incoherent oscillations of the ALP field over the total monitoring time-span ($\Delta\,t_{\rm mon}$) 
of all epochs of observations. 

For the case of \emph{incoherent oscillations}, we cannot expect the phases $\delta_\text{em}^{(n)}$ to be related in any simple manner and thus we have no handle on the second sinusoidal factor in eq.~(\ref{Eq:DiffBir_multi_incoherent}), which thus should be averaged and we would combine all epochs of observations by means of weighted averages, as further detailed in section~\ref{sec:WtAvg}.

However, ALP field oscillations are \emph{coherent} over a time-scale $\tau_{\rm c} \approx 2/m_a v^2$, where, $v \sim 10^{-3}$ is the typical velocity dispersion of an ALP dark matter halo \cite{Marsh2016}. Therefore,
for multiple observations conducted over $\Delta\,t_{\rm mon}$ such that,
\begin{equation}
\label{eq:tau_coherence}
\Delta\,t_\text{mon} \ll \tau_{\rm c} = 42\,{\rm yr}\,\left(\frac{10^{-18}\,{\rm eV}}{m_a}\right)\left(\frac{10^{-3}}{v}\right)^{2},
\end{equation}
the ALP field is well-described by oscillation having a single amplitude and phase. 
Here, $\Delta\,t_{\rm mon} = \Delta\,t_{\rm o,mon}/(1 + z_\mathrm{qso})$
for observations time-span $\Delta\,t_{\rm o,mon}$.
When the 
condition above is satisfied for a monitoring campaign, only a single phase, 
$\delta_\mathrm{em}^\mathrm{ref}$, for an arbitrary reference epoch, $t^\mathrm{ref}_\mathrm{em}$, 
needs to be
considered, and we can write 
\begin{equation} \label{Eq:DiffBir_multi}
    \Delta \theta_{a,\text{lens}}^{\,(n)} = K \sin \left[ \frac{m_a\, \Delta t_\text{d}}{2} \right] \sin \left[ m_a\, \Delta t_\text{em}^{(n)}  + {\delta}_\text{em} - \frac{\pi}{2} \right],
\end{equation}
where, $\Delta t_\text{em}^{(n)} = t_\text{em}^{(n)} - t_\text{em}^\text{ref}$ is the time elapsed between the reference epoch and the $n^\text{th}$ epoch and $\delta_\mathrm{em} =  m_a t^\mathrm{ref}_\mathrm{em} + \delta^\mathrm{ref}_\mathrm{em}$ is the shifted-phase compared to that of the reference epoch.
As all the above quantities are specified in emission frame, 
$\Delta t_\text{em}^{(n)} = \Delta t_\text{o}^{(n)} / (1+z_\text{qso})$. 

Thus, for coherent oscillations, a model for the observations of $N$ epochs has three 
parameters, $K, m_a$ and the effective 
phase $\delta_\mathrm{em}$, and they
can be measured when a source is observed for $N > 3$.
In section~\ref{sec:ProfileLikelihood}, we account for this condition, and present a method based on profile likelihoods to obtain upper limits on $K$ as a function of $m_a$ and convert them into upper limits on $\gag$ as a function of $m_a$.
Since the signal is a product of two sinusoidal oscillations, 
different lensing systems
would give rise to different time-dependent modulation of $\dthetalens$ depending 
on $\Delta\,t_{\rm d}$.
This implies, a discovery in one system would make a precise prediction for how the same ALP signal should look like in another lens system, allowing for a 
consistency check that only 
few other astrophysical ALP search methods allow.

\section{Observations and data reduction of CLASS B1152+199} \label{sec:Obs}

\begin{figure}
    \centering
    \includegraphics[width=0.7\linewidth]{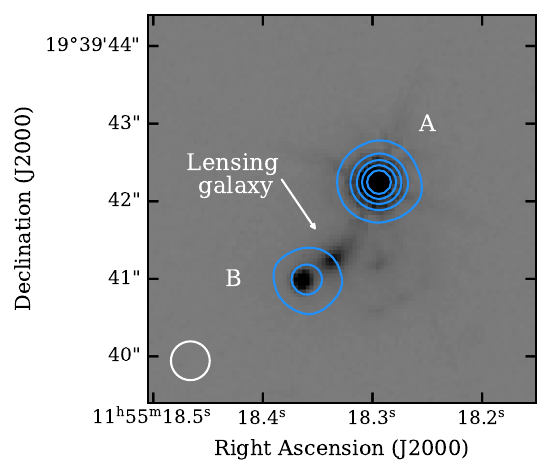}
    \caption{ 
    Hubble Space Telescope image of CLASS\,B1152+199 observed using the F814W filter is overlaid by the VLA total intensity radio contour at 4.5\,GHz. The resolution of the radio image is 0.5\,arcsec shown as the white circle, and the blue contours are at 2, 10, 18, 26 and 34\,$\rm mJy\,beam^{-1}$. The two lensed images A and B are observed in both radio and optical frequencies whereas the lensing galaxy is only visible in the optical observations.}
    \label{fig:intIm}
\end{figure}

We monitored the strong gravitational lens system CLASS\,B1152+199 \cite{Myers1999}, henceforth referred to as B1152+199,
using the VLA in A-configura\-tion.
In B1152+199, a background linearly polarised quasar at redshift $z_{\rm qso}=1.019$ is strongly lensed by a foreground starforming galaxy at redshift $z_{\rm gal}=0.439$ 
into two lensed images separated by $1.56^{\prime\prime}$. 
It was first observed during phase 3 of the Cosmic Lens All-Sky Survey (CLASS) in 1998 
with the VLA in its A-configuration at
8.46 and 14.94\,GHz \cite{Myers1999}. They found identical spectra for the two lensed images and established B1152+199
to be a strongly lensed system. The ratio of flux densities observed in the images was $\sim$ 3:1. In this paper, we will denote the brighter and the fainter lensed image components of B1152+199 as `A' and `B', respectively, see figure~\ref{fig:intIm}. It has been observed
sporadically over 20\,years. Previous observations aimed to either model the lens system \cite{Rusin2002}, determine time delay to constrain 
the Hubble's constant $H_0$ \cite{Munoz2001}, or study the magnetic fields in the interstellar medium of the lensing galaxy \cite{Mao2017}.

\begin{table}[t]
\centering
\begin{tabular}{ccccc}
\toprule
$n$ & Epochs & MJD  & $\Delta t_\mathrm{em}$ [day] & Reference \\
\midrule
1 & 08 March 2022 & 59646.247384 & 0 & This work \\
2 & 28 March 2022 & 59666.179931 & \qquad 9.872485 & This work \\
3 & 17 April 2022 & 59686.137292 & \qquad 19.757260 & This work \\
4 & 07 May 2022 & 59706.088194 & \qquad 29.638836 & This work \\
5 & 27 May 2022 & 59726.072546 & \qquad 39.536980 & This work\\ [0.2cm]
6 & 13 November 2012 & 56244.544825 & $-1684.845250$ & Mao et al.~\cite{Mao2017}\\
\bottomrule
\end{tabular}
\caption{Epochs of observations of B1152+199 taken with the VLA used for searching for ALPs in this work. The five epochs labelled 1--5 were analysed for this work, and their combined analyses are referred to as `5-Epochs'. The combined analyses performed by combining `5-Epochs' along with the archival data at epoch 6 is referred to as `6-Epochs'. For the purpose of computing time intervals, we use modified Julian date (MJD). 
Time intervals in the emission frame ($\Delta\,t_{\rm em}$) with respect to the reference epoch 1 are listed in fourth column.}
\label{tab:epochs}
\end{table}

\begin{table}[t]
\centering
\begin{tabular}{rccc}
\toprule
Object observed & Band & Observing duration & Purpose \\
(in sequence)   &      &  [min] & \\
\midrule
3C\,286 & L, S, C & 6, 6, 6 & Absolute flux density, frequency\\
        &         &         & and polarisation calibration\\
J1158+2450 & C & 5 & Time-dependent gain calibration\\
B1152+199 & C & 10 & Target \\
J1158+2450 & C, L & 2, 2 & Time-dependent gain calibration \\
B1152+199 & L & 11.5 & Target \\
J1158+2450 & L, S & 2, 2 & Time-dependent gain calibration \\
B1152+199 & S & 5 & Target \\
J1158+2450 & S & 2 & Time-dependent gain calibration \\
J1407+2827 & L, S, C & 10, 8, 8 & On-axis leakage calibration \\
\bottomrule
\end{tabular}
\caption{Summary of the observations taken with the VLA for the 5-Epochs listed in table~\ref{tab:epochs}. The rows are listed in time-order from the start of observing session, and the bands in each row are also in time-order.
}
\label{tab:Obs}
\end{table}

Our observations of B1152+199 were conducted over five epochs spanning three months between March and May 2022, with a cadence of roughly 20\,days. 
Table~\ref{tab:epochs} lists the epochs of observations used for this study and in table~\ref{tab:Obs} the observations strategy is summarised.
Full Stokes data were recorded in the 1 to 8 GHz frequency range, covered using three receiver systems of the VLA, namely, L-band (1--2\,GHz), S-band (2--4\,GHz) and C-band (4--8\,GHz).
For all epochs, B1152+199 was sequentially observed at each of the three bands by switching the receivers, starting with 10\,min at C-band, 11.5\,min at L-band, and 5\,min at S-band.
The scans on B1152+199 at each of the bands were interleaved with scans on a secondary calibrator, J1158+2450. The primary and polarisation angle calibrator, 3C\,286, and a weakly polarised leakage calibrator, J1407+2827, were observed at the beginning and end, respectively, at each epoch and at all three frequency bands. 

These data were analysed using the \texttt{NRAO\,CASA} package \texttt{version\,6.2.1.7} \cite{casa}, where full polarisation calibration and spectro-polarimetric imaging of B1152+199 were performed. We followed the standard spectro-polarimetric data analyses protocol, wherein the data were first flagged to mitigate bad data arising from radio frequency interference and telescope operational factors. The secondary calibrator, J1158+2450, was used to solve the time dependent complex gain, and the primary calibrator, 3C\,286, was used to solve frequency dependent telescope gains (bandpass). To solve for the on-axis instrumental leakage, the weakly polarised source, J1407+2827 was used. 3C\,286 was also used to calibrate the absolute polarisation angle. While the data at each of the frequency bands were independently calibrated, we ensured smooth continuity of solutions across the bands by jointly solving the absolute total flux density scale for 3C\,286 using the Perley \& Butler model \cite{Perley&Butler2017} across the entire 1 to 8\,GHz range. After solving all the antenna based gains, the solutions were transferred to the target source, B1152+199.

Since we present combined analyses of B1152+199 at all the three bands at each epoch, the effective time duration spent was 36.2\,min 
including overheads with observing J1158+2450 and telescope slewing. Thus, the total duration over which our data are effectively averaged is $\Delta\,t_{\rm avg} = 17.93$\,min in the rest-frame of the lensed quasar. Furthermore, we also include the measurement of $\dthetalensS$ \cite{Basu2021} observed on 13 November 2012 (MJD 56244.544825) \cite{Mao2017} in our analysis.\footnote{We will refer to combined analyses of epochs 1 through 5 (table~\ref{tab:epochs}) with the data observed for this work as `5-Epochs' and with all the six epochs as `6-Epochs'.} 
Equating $\Delta\,t_{\rm avg}$ with the oscillation period of the ALP field determines the maximum $m_a \approx 3.8 \times 10^{-18}$\,eV 
to which the observations of a single epoch are sensitive to \citep{Basu2021}. The time delay between the two lensed images in B1152+199, $\Delta t_{\rm d} = 13.3$\,day \cite{Rusin2002,Basu2021}, 
provides an estimate of the minimum 
$m_a \approx 3.6 \times 10^{-21}$\,eV
for each individual epoch. The overall duration of our monitoring in the frame of the quasar, $\Delta t_{\rm mon} = 39.54$\,day for 5-Epochs,
and $\Delta t_{\rm mon} = 1724.65$\,day for 
6-Epochs, 
determines the minimum
$m_a$ of $1.2 \times 10^{-21}$\,eV and 
$2.8 \times 10^{-23}$\,eV,
respectively, for the combined analyses of 
the data.

\subsection{Imaging and Stokes $I$ spectrum} \label{sec:totI}

\begin{figure}[t]
    \centering
    \includegraphics[width=\linewidth]{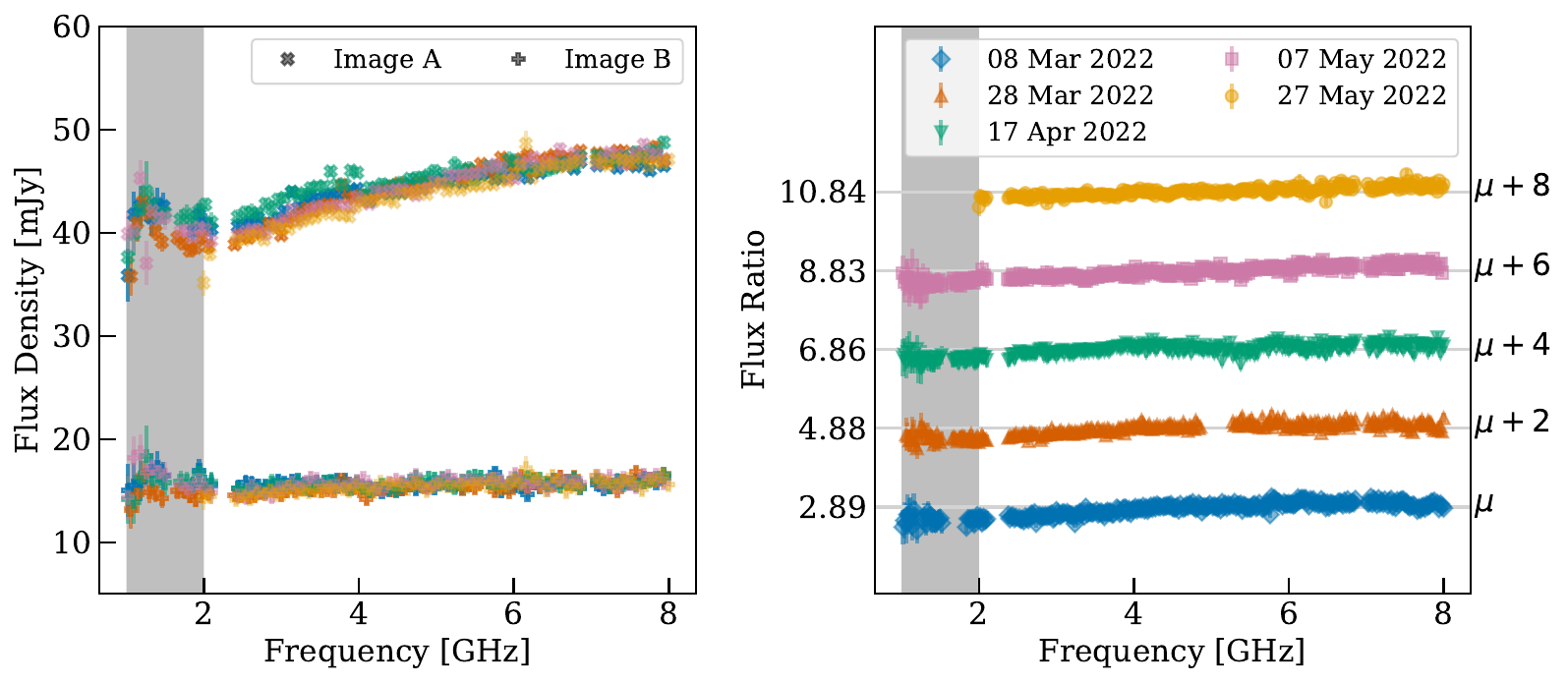}
    \caption{\textit{Left panel} shows the Stokes\,$I$ spectra averaged over 20\,MHz for the lensed images A and B in B1152+199. At all the epochs, the Stokes\,$I$ flux densities are within the 5\% flux scale uncertainty of the VLA. \textit{Right panel} shows the lens magnification factor, $\mu = S_{\rm A}(\nu)/S_{\rm B}(\nu)$, of the lensed images as a function of frequency for the five epochs. To avoid overlap, each epoch is plotted with an offset shown in the right-hand $y$-axis. The grey lines are the corresponding best-fit $\mu$, again plotted with an offset. The shaded region in both the panels covering 1--2\,GHz shows significant fluctuations due to blending of image A and B and the data in that frequency range were not used in our analyses.}
    \label{fig:Spectrum}
\end{figure}

The full Stokes calibrated data of B1152+199 were deconvolved using the \texttt{CLEAN} algorithm to produce images of the Stokes\,$I$, $Q$ and $U$ parameters averaged over 20\,MHz frequency channels covering the 1--8\,GHz frequency range of our observations. Figure~\ref{fig:intIm} shows a typical Stokes\,$I$ contour map of B1152+199 at 4.5\,GHz observed on 08\,March\,2022  overlaid on the optical image obtained with the Hubble Space Telescope. Since the lensed image components A and B are separated by $1.56^{\prime\prime}$, we enforced an angular resolution of $0.5^{\prime\prime}\times0.5^{\prime\prime}$ for the channel images above 2\,GHz in order to resolve them. However, the angular resolution $\gtrsim1^{\prime\prime}$ provided by the VLA in the 1--2\,GHz range leads to significant blending of the lensed images, making it difficult to reliably extract the flux densities of the lensed image components. This is seen clearly in the shaded region of figure~\ref{fig:Spectrum} where the flux densities and lens magnification below 2\,GHz show significant fluctuations. We have therefore, restricted our analyses to the frequency range 2--8\,GHz, using 300 channels with a channel-width of 20\,MHz.

\begin{table}
\centering
\begin{tabular}{cccc}
\toprule
Epoch & $S_{\rm 0,A}$ & $\alpha$ & $\mu$   \\
  $(n)$     & [mJy]  &   &   \\
\midrule
$1$ & $44.39\pm0.03$ & $0.107\pm0.002$ & $2.900\pm0.005$ \\
$2$ & $44.28\pm0.03$ & $0.155\pm0.002$ & $2.904\pm0.006$ \\
3 & $45.13\pm0.04$ & $0.089\pm0.003$ & $2.877\pm0.008$ \\
$4$ & $44.14\pm0.03$ & $0.148\pm0.002$ & $2.849\pm0.006$ \\
$5$ & $42.16\pm0.07$ & $0.142\pm0.004$ & $2.794\pm0.014$ \\
\bottomrule
\end{tabular}
\caption{Flux density of image A $S_{\rm 0,A}$ [mJy] at 4.5 GHz, the spectral index $\alpha$ and magnification ratio $\mu$ of image A with respect to image B computed by modelling the total intensity spectrum as a power-law.}
\label{tab:flux}
\end{table}

The typical root mean square (rms) noise in the 20\,MHz channel images 
ranges between $150$ and $300\, \rm \upmu Jy\,beam^{-1}$ for all five epochs of observation. These noise levels are consistent with the theoretical expectations for the VLA within a factor of 2. The flux density at each of the 20\,MHz channels for Stokes\,$I$, $Q$ and $U$ were obtained by modelling the lensed components A and B in B1152+199 as unresolved 2-D Gaussian. Stokes\,$I$ spectrum of A and B are shown in the left-hand panel of figure~\ref{fig:Spectrum}. The measured flux densities in each of the five epochs are within the $\sim 5\%$ flux-scale accuracy of the VLA \cite{Perley&Butler2017}, and our observations do not show indication of temporal variation of Stokes\,$I$ in either of the two lensed images. 

The Stokes\,$I$ spectra for images\,A and B are both well described by a power-law $S(\nu) = S_0\,(\nu/\nu_0)^{\alpha}$, where, $S(\nu)$ is the flux density at a frequency $\nu$, $\alpha$ is the spectral index, and $S_0$ is the flux density at a pivot frequency $\nu_0$ chosen at 4.5\,GHz. The spectra for the two images were fitted by keeping the same $\alpha$ for both and allowing for the flux density to scale as $S_{\rm B}(\nu) = S_{\rm A}(\nu)/\mu$, where $\mu$ is the lens magnification ratio, and $S_{\rm A}$ and $S_{\rm B}$ are the flux densities of image\,A and B. 
The Stokes\,$I$ spectra for A and B were simultaneously fitted with three free-parameters $S_{\rm 0,A}$, $\mu$ and $\alpha$, and we did not invoke the additional systematic error on the flux-scale.\footnote{We also tested the robustness of our fitting approach by independently fitting for image A and B. Except for increased error, we did not find any significant differences in the values of $S_{\rm 0,A}$ and $S_{\rm 0,B}$, and thereby on $\mu$ and $\alpha$.} The best-fit parameters are summarised in table~\ref{tab:flux}, and within the uncertainties, we do not find any significant variation across the five epochs. From our data, the weighted mean $\alpha$ over the five epochs is found to be $\langle \alpha\rangle = 0.13 \pm 0.03$, which is typically expected for quasars. The weighted mean magnification ratio $\langle \mu\rangle = 2.88 \pm 0.03$ is consistent with previous narrow-band measurements for B1152+199 \cite{Myers1999, Edwards2001}.

\begin{figure}[t]
    \centering
    \includegraphics[width=0.7\linewidth]{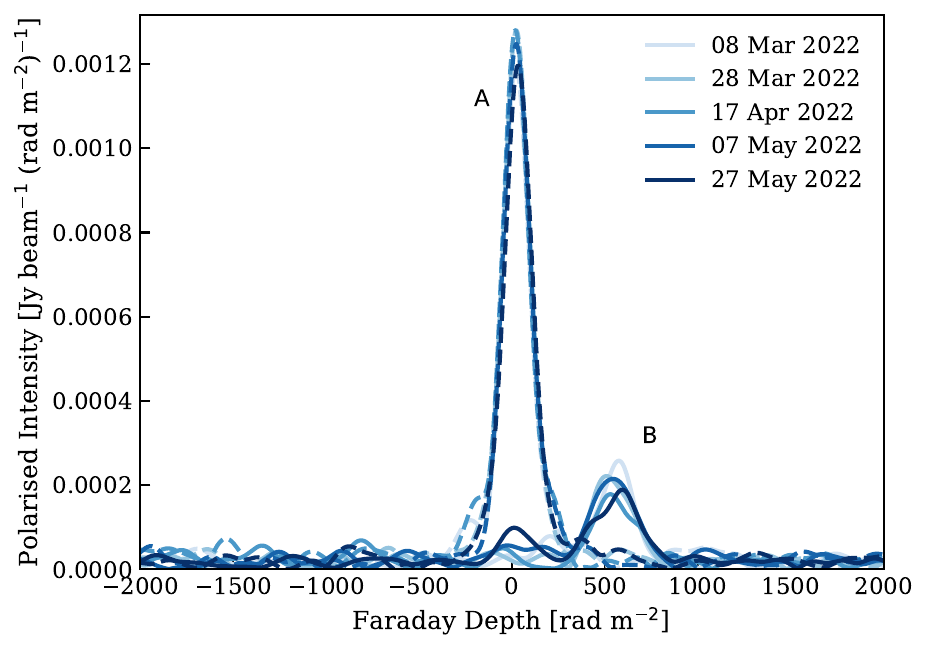}
    \caption{Faraday depth spectra of the two lensed images in B1152+199 for the 5-Epochs obtained from RM synthesis. The two peaks correspond to Image\,A (dashed lines) and B (solid lines).}
    \label{Fig:RMspectrum}
\end{figure}

\subsection{Polarisation analysis and birefringence angle} \label{sec:PolAna}

We did not find any indication of time variability in the Stokes\,$I$ emission of images A and B over the five epochs. This suggests that any intrinsic variability of the lensed quasar in B1152+199 from astrophysical effects are unlikely. Here we 
analyse the linear polarisation parameters Stokes\,$Q$ and $U$, and estimate differential birefringence, $\dthetalens$. The observed orientation ($\theta_{\rm obs}$) of the electric vector of linear polarisation at a frequency $\nu$ is given by,
\begin{equation}
    \theta_{\rm obs}(\nu) = \frac{1}{2} \arctan \left[ \frac{U(\nu)}{Q(\nu)} \right].
    \label{eq:PA}
\end{equation}
$\theta_{\rm obs}$ depends on the frequency because the polarised light also undergoes \textit{chromatic} Faraday rotation as it traverses through magneto-ionic medium lying in between the emitting source and the observer. The intrinsic angle of polarisation ($\theta_0$) at emission is related to $\theta_{\rm obs}$ as,
\begin{equation}
    \theta_{\rm obs}(\nu) = \theta_0 + {\rm RM}\,\left(\frac {c} {\nu}\right)^2.
    \label{eq:RM}
\end{equation}
Here, RM is the Faraday rotation measure which depends on the integral of the magnetic field component along the line of sight weighted by the number density of free-electrons. As the polarised light of the two lensed images of the quasar in B1152+199 traverses through different paths in the lensing galaxy, and thereby different magneto-ionic media, the components A and B undergo different amount of Faraday rotation characterised by $\rm RM_A$ and $\rm RM_B$.

We implement the technique of RM synthesis \cite{Burn1966, Brentjens&Bruyn2005} to determine $\rm RM_A$ and $\rm RM_B$ using the \texttt{pyrmsynth}\footnote{\url{https://github.com/mrbell/pyrmsynth/tree/master}} package. For each epoch and each image\,A and B, Stokes\,$Q$ and $U$ spectra were extracted for each of the 300 20-MHz channels covering 2--8\,GHz as described in section~\ref{sec:totI}. Since the signal to noise ratio of Stokes\,$Q$ and $U$ images per 20\,MHz channel are low, the Stokes\,$Q$ and $U$ flux densities were fitted by fixing the positions of image\,A and B to their corresponding Stokes\,$I$ positions. The Stokes $Q$ and $U$ spectra extracted from 5-Epochs are presented in  figure~\ref{fig:QUspec} of appendix~\ref{app:QUspec} as the data points along with, for comparison, the modelled Stokes\,$Q,U$ spectra using data at epoch $n=6$ \cite{Mao2017} as the lines. Within the errors, our data are in good agreement with the modelled Stokes\,$Q$ and $U$ spectra for all the epochs, suggesting that the data calibration systematics across the six epochs are low. These Stokes\,$Q$ and $U$ spectra were then used to perform RM synthesis for Image\,A and B at each epoch independently. The Faraday depth spectra, polarised intensity as a function of Faraday depth,\footnote{In the context of RM synthesis, Faraday depth is defined as the RM at which polarised emission accumulates from different physical depth along the line of sight.} obtained from RM synthesis are presented in figure\,\ref{Fig:RMspectrum}, and $\rm RM_A$ and $\rm RM_B$ are given by the location of their respective peaks. The RM values were found to be $\sim \rm +20\, rad\,m^{-2}$ for Image\,A and $\sim \rm +550\, rad\,m^{-2}$ for Image\,B. Note that, RM synthesis provides Stokes\,$Q$ and $U$, and thereby, $\theta_{\rm obs}$ at an effective frequency ($\nu_{\rm eff}$) which 
corresponds to the channel-noise weighted average of the $\lambda^2$-coverage of the observations \cite{Brentjens&Bruyn2005}. Table~\ref{tab:polarization_AB} lists $\nu_{\rm eff}$ and Stokes\,$Q$ and $U$ parameters.

To mitigate the effect of Faraday rotation, and subsequently compute $\dthetalensS$ using eq.~(\ref{eq:difftheta}), we compute the frequency-independent, intrinsic angle of polarisations, $\theta_{\rm 0,A}$ and $\theta_{\rm 0,B}$ from eqs.~(\ref{eq:PA}) and (\ref{eq:RM}) as,
\begin{equation}
    \theta_{\rm 0,A} = \frac 1 2 \arctan\left[\frac{U_{\rm A}(\nu_{\rm eff})}{Q_{\rm A}(\nu_{\rm eff})}\right] - {\rm RM_A}\left(\frac{c}{\nu_{\rm eff}} \right)^2 \textrm{\ and\ } \theta_{\rm 0,B} = \frac 1 2 \arctan\left[\frac{U_{\rm B}(\nu_{\rm eff})}{Q_{\rm B}(\nu_{\rm eff})}\right] - {\rm RM_B}\left(\frac{c}{\nu_{\rm eff}} \right)^2\!\!.
    \label{eq:theta0}
\end{equation}
$\theta_{\rm 0,A}$ and $\theta_{\rm 0,B}$ for each epoch were then used to compute $\Delta\,\theta_{a,{\rm lens}}^{(n)}$ and derive constrains on $\gag$ as a function of $m_a$. In table~\ref{tab:rm_theta_AB}, we list the values of $\rm RM_A$ and $\rm RM_B$, the values of polarisation angle at $\nu_{\rm eff}$.

\begin{table}
\centering
\begin{tabular}{cccccc}
\hline
Epoch
& $\nu_\text{eff}$
& $Q_{\mathrm{A}}(\nu_\text{eff})$
& $U_{\mathrm{A}}(\nu_\text{eff})$
& $Q_{\mathrm{B}}(\nu_\text{eff})$
& $U_{\mathrm{B}}(\nu_\text{eff})$ \\
$(n)$ & [GHz]
& [mJy\,beam$^{-1}$]
& [mJy\,beam$^{-1}$]
& [mJy\,beam$^{-1}$]
& [mJy\,beam$^{-1}$] \\
\hline
1 & $4.343$ & $0.80 \pm 0.02$ & $0.89 \pm 0.02$
  & $0.23 \pm 0.03$ & $-0.113 \pm 0.020$ \\

2 & $3.854$ & $0.76 \pm 0.03$ & $1.03 \pm 0.02$
  & $0.16 \pm 0.04$ & \quad$0.146 \pm 0.035$ \\

3 & $4.193$ & $0.74 \pm 0.02$ & $1.04 \pm 0.01$
  & $0.17 \pm 0.04$ & $-0.002 \pm 0.012$ \\

4 & $4.016$ & $0.74 \pm 0.02$ & $1.01 \pm 0.02$
  & $0.20 \pm 0.04$ & \quad$0.088 \pm 0.023$ \\

5 & $4.131$ & $0.79 \pm 0.02$ & $0.90 \pm 0.01$
  & $0.17 \pm 0.05$ & \quad$0.049 \pm 0.017$ \\
\hline
\end{tabular}
\caption{Stokes\,$Q$ and $U$ parameters at $\nu_{\rm eff}$ for Image~A and B in B1152+199 obtained from the technique of RM synthesis. The uncertainties correspond to $1\sigma$ values.}
\label{tab:polarization_AB}
\end{table}

\begin{table}
\centering
\begin{tabular}{cccccc}
\hline
Epoch
& RM$_{\mathrm{A}}$
& RM$_{\mathrm{B}}$
& $\theta_A(\nu_\text{eff})$
& $\theta_B(\nu_\text{eff})$
& $\Delta \theta_{a,\text{lens}}^{(n)}$ \\
$(n)$& [rad\,m$^{-2}$]
& [rad\,m$^{-2}$]
& [deg]
& [deg]
& [deg] \\
\hline
1 & $21.3 \pm 1.8$ & $575.0 \pm \;9.4$  & $23.9 \pm 0.4$ & $-13.3 \pm 2.4$  & $\quad8.44_{-3.59}^{+3.58}$ \\

2 & $22.7 \pm 2.0$    & $517.5 \pm 24.3$ & $26.7 \pm 0.5$   & $\quad21.4 \pm 4.9$ & $-3.20_{-9.84}^{+9.68}$ \\

3 & $18.7 \pm 1.3$ & $533.7 \pm 14.2$  & $27.3 \pm 0.4$ & $\;-0.3 \pm 2.1$   & $-1.61_{-4.61}^{+4.75}$ \\

4 & $23.3 \pm 1.7$ & $539.1 \pm 28.4$  & $26.8 \pm 0.5$ & $\quad12.0 \pm 4.0$  & $-0.63_{-9.93}^{+9.64}$ \\

5 & $32.6 \pm 1.9$ & $591.9 \pm 19.1$  & $24.4 \pm 0.4$ & $\quad8.0 \pm 3.8$   & $\quad4.88_{-7.05}^{+6.60}$ \\
\hline
\end{tabular}
\caption{Rotation measure obtained from RM synthesis for lensed images A ($\rm RM_A$) and B ($\rm RM_B$), and the observed polarisation angles, $\theta_{\rm A}$ and $\theta_{\rm B}$ at $\nu_\text{eff}$ in B1152+199.
The last column is for the differential birefringence angle ($\dthetalens$) where we quote the median value obtained from the distributions in figure~\ref{fig:DiffBir_dist}. The uncertainties correspond to $1\sigma$ values and to the 68\% confidence interval for $\dthetalens$.}
\label{tab:rm_theta_AB}
\end{table}

\begin{figure}[t]
    \centering
    \includegraphics[width=0.9\linewidth]{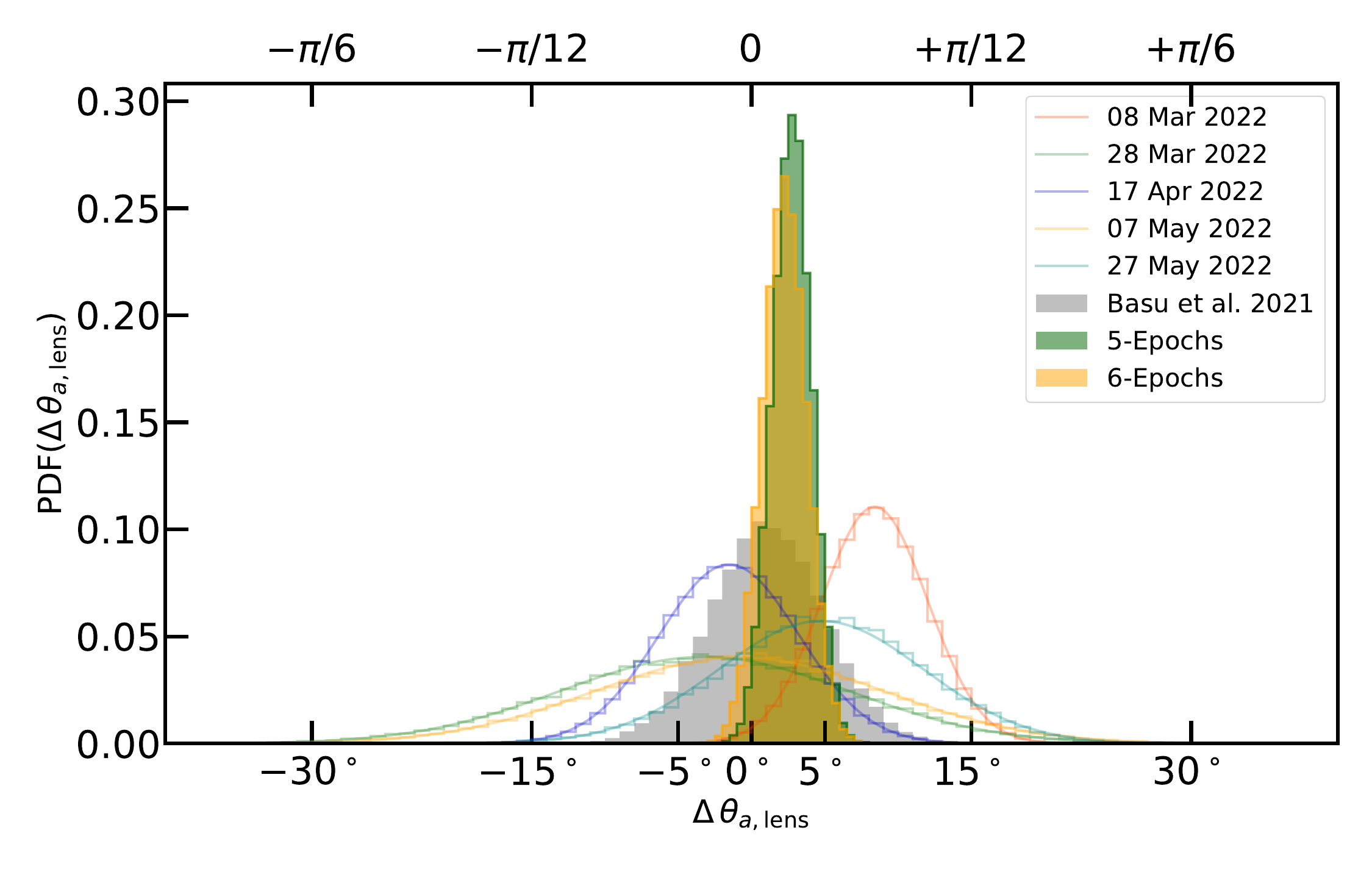}
    \caption{The PDFs of $\Delta \theta_{a,\mathrm{lens}}$ obtained from the 5-Epochs observation of B1152+199, drawn from a sample of 50,000 relisations are well-approximated by the Gaussian distribution shown as solid lines. The shaded green and orange histograms are for 5-Epochs and 6-Epochs combined $\dthetalens$ through weighted average discussed in section~\ref{sec:WtAvg}. For comparison, the distribution of $\dthetalens$ obtained for a single epoch in Basu et al. \cite{Basu2021} is shown as the shaded grey histograms.}
    \label{fig:DiffBir_dist}
\end{figure}

Note that, in order to derive the error on the quantities that are used to estimate $\dthetalens$, we performed Monte-Carlo simulation from random realisations of the data. Since, quantities estimated through polarisation analysis are non-linearly related, determining robust error on $\dthetalens$ is important. A detailed description of the method of error estimation is provided in appendix~\ref{app:error}. The distribution of $\dthetalens$ from these random realisations at each epoch are shown in figure~\ref{fig:DiffBir_dist} as the unfilled histograms. In table~\ref{tab:rm_theta_AB}, we report the median $\dthetalens$ and 68\% interval around the median obtained from the distribution of the random realisations at each of the epochs.

\section{Constraints on birefringence angle} \label{sec:constrain}

In this section, we compute constraints on the differential birefringence angle ($\dthetalensS$) by combining the multi-epoch measurements by considering two scenarios. 
In section~\ref{sec:ProfileLikelihood}, we  performed a detailed analysis by considering that the ALP field strength oscillation, and thereby $\dthetalens$ obtained in table~\ref{tab:rm_theta_AB}, are coherent over $\tau_c$, which is significantly larger than the span of our monitoring observations.
Then in section~\ref{sec:WtAvg}, we statistically combine $\dthetalens$ as a weighted mean. This is done to compare with a single epoch observation of B1152+199 on 13\,November\,2012 \cite{Mao2017} (denoted as epoch 6 in table~\ref{tab:epochs}) which was used in \cite{Basu2021} to constrain $\gag$ as a function of $m_a$. 
They reported $\Delta \theta_{a,\mathrm{lens}}^{(6)} = 1.04^{+3.90}_{-1.80}$\,degree (68\% confidence interval), corresponding to a 95\% upper limit $|\Delta \theta_{a,\mathrm{lens}}^{(6)}| < 8.71$\,degree.

\subsection{Profile likelihood ratio approach} \label{sec:ProfileLikelihood}

Here we combine multiple $\Delta\,\theta_{a,{\rm lens}}^{(n)}$ by considering the effect of ALP field oscillation 
discussed in section~\ref{Sec:MultiEpoch}, wherein, $\Delta\,\theta_{a,{\rm lens}}^{(n)}$ is related to a reference epoch given by eq.~(\ref{Eq:DiffBir_multi}). That is, the condition for coherent ALP field oscillation
in eq.~(\ref{eq:tau_coherence}) is satisfied over the entire time span of our study.
For this, we construct the Gaussian likelihood 
\begin{equation} \label{Eq:LikelihoodFn}
    \mathcal{L}(K, m_a, \delta_\text{em}; \dthetalens) = \prod_{n=1}^N \frac{1}{\sigma_n\,\sqrt{2\pi}} \exp{\left[ -\frac{1}{2\sigma_n^2} \left(\Delta \theta_{a,\text{lens}}^{(n)} 
    - \Delta \theta_{a,\text{lens}}^{(n),\text{mod}} \right)^2 \right]} \;,
\end{equation}
where, $N$ is the number of epochs and $\sigma_n$ is the 68\% confidence on $\Delta \theta_{a,\text{lens}}^{(n)}$, and $\Delta\,\theta_{a,{\rm lens}}^{(n),{\rm mod}} = \Delta\,\theta_{a,{\rm lens}}^{(n),{\rm mod}}(K,m_a,\delta_\mathrm{em})$ is the model given by eq.~(\ref{Eq:DiffBir_multi}). 
Note that, from definition in eq.\,(\ref{Eq:K}), $K$ is a positive quantity, and since, $|\Delta \theta_{a,\text{lens}}| < 180^\circ$, $K< 180^\circ$.
The phase at emission, $\delta_{\rm em}$, can be anywhere between 0 and $2\pi$.
Thus, we use the following constraints,
\begin{gather}
    0^\circ \le K \le 180^\circ \;, \label{eq:Kcondition}\\
    0 \le \delta_\text{em} \le 2\,\pi \;, \label{eq:delta}\\
    \frac{1}{2\,\tau_{\rm c}} \lesssim m_a \lesssim \frac{2\pi}{\Delta\,t_\text{avg}} \;. \label{eq:mass}
\end{gather}
The last condition accounts for the fact that $\Delta\,\theta_{a,{\rm lens}}^{(n)}$ are related such 
that we can use eq.~(\ref{Eq:DiffBir_multi}) to model the effect of oscillating ALP field at emission in our observations. 
For the purpose of our analysis, we consider 08 March 2022 (MJD 59646.247384) as the reference epoch, $t_{\rm em}^{\rm ref}$, such that, $\Delta t_{\rm em}^{(1)} = 0$ for $\Delta\,\theta_{a,{\rm lens}}^{(1)}$.

\begin{figure}
    \centering
    \includegraphics[width=\linewidth]{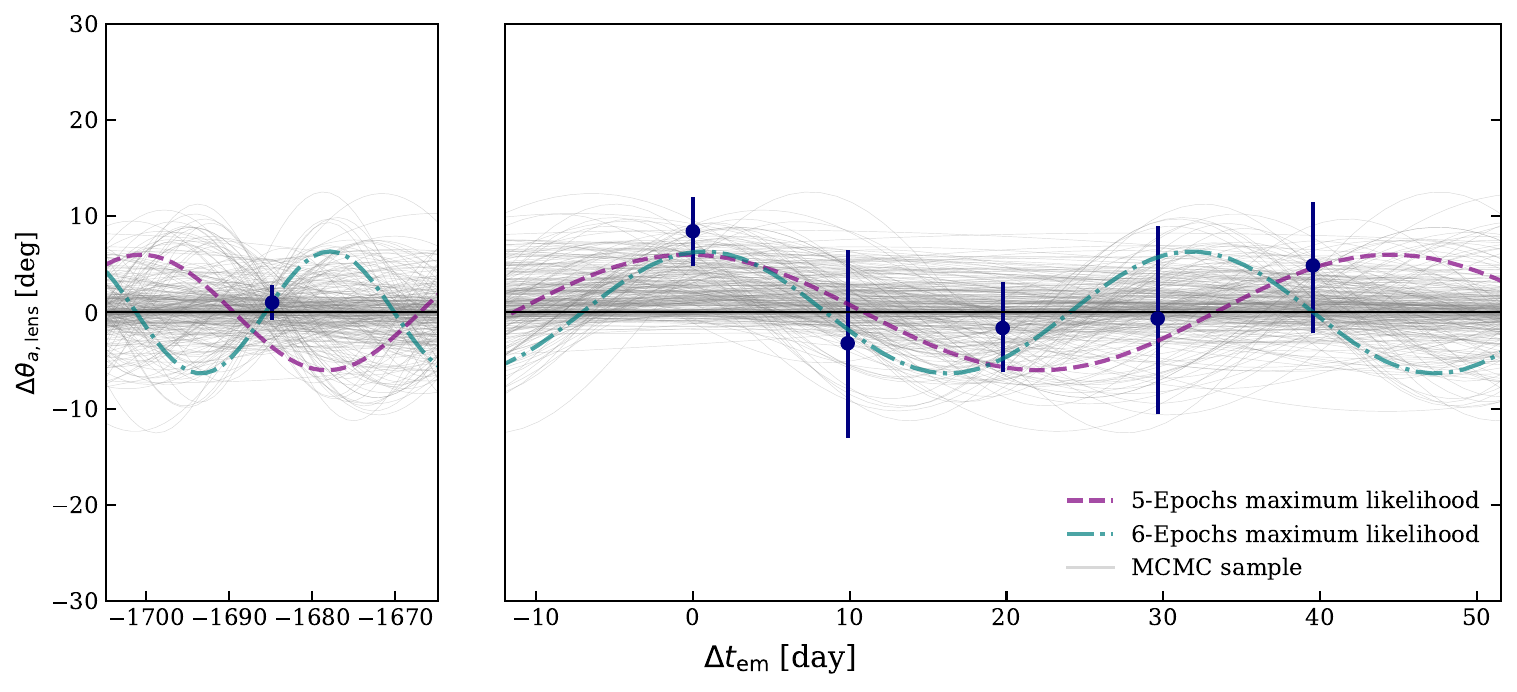}    
    \caption{Estimated differential birefringence angle, $\Delta\,\theta_{a,{\rm lens}}^{(n)}$, from B1152+199 as a function of time at emission, $\Delta\,t_{\rm em}$, where the observations on 08 March 2022 is assumed as the reference, are shown as the data points. The dashed magenta and dash-dotted turquoise lines show the maximum likelihood estimates
    for 5-Epochs and 6-Epochs, respectively. 
    The grey lines are randomly picked 300 MCMC sample produced for the 6-Epochs. 
    The corresponding posterior distributions are shown in Fig.~\ref{fig:CornerPlot}. The black horizontal line represents the null-hypothesis.
    }
    \label{fig:DiffBir_sinFit}
\end{figure}

Due to the oscillatory nature of the model, we expect multi-modality in the posterior distribution of model parameters. 
Because of roughly equal cadence of observations,
any oscillating solution corresponding to that associated frequency and its harmonics are possible. Hence, in order to explore the possibility of other non-trivial solutions, we constrained 
$m_a < 2\,\pi/(3\, t_{\rm cad})$ where $t_{\rm cad} \approx 10$\,days is the cadence of observations in rest frame instead of eq.~(\ref{eq:mass}).
The posterior distribution of the parameters are computed using Markov Chain Monte Carlo (MCMC) sampling 
(see appendix~\ref{app:Posterior}). 
Figure~\ref{fig:DiffBir_sinFit} shows the MCMC samples as the grey lines and the maximum likelihood of 5- and 6-Epochs are shown as the magenta dashed line and turquoise dash-dotted line, respectively. 
The posterior distribution of $m_a$ exhibit two high density regions (see figure~\ref{fig:CornerPlot}), and the maximum likelihood estimate (MLE) lie in the second, sub-dominant, peak at higher $m_a$. In order to determine confidence intervals, we decomposed the posterior distribution of $m_a$ by fitting two Gaussians and use the 68\% interval around the maximum likelihood value. 
For 5-Epochs, we obtain $K=7.47$\,degree [(3.13, 11.81)\,degree; 68\% interval] and $m_a= 1.07 \times 10^{-21}$\,eV [$(0.71, 1.53) \times 10^{-21}$\,eV; 68\% interval]. For 6-Epochs, we obtain $K=6.51$\,degree [(4.14, 8.88)\,degree; 68\% interval] and $m_a = 1.54 \times 10^{-21}$\,eV [$(0.54, 1.60) \times 10^{-21}$\,eV; 68\% interval].
Although the MLEs correspond to an oscillatory model, we emphasise that the majority of the MCMC samples support the null hypothesis seen as a higher density of samples close to $\dthetalensS=0$\,degree in figure~\ref{fig:DiffBir_sinFit}.

We now compare the ALP-induced birefringence model to the null hypothesis where no such oscillatory signal is present, i.e., 
$K=0$ for the model in eq.~(\ref{Eq:DiffBir_multi}). For the null hypothesis, we find a $\chi_{\rm red}^2 = 1.1$ ($\chi^2$ per degrees of freedom) over all the six
measurements of $\dthetalens$, i.e., our data are consistent with the null hypothesis with a $p$-value of 0.38. In comparison, for the 6-Epochs best-fit model above, $\chi^2_{\rm red} = 0.59$ with a $p$-value of 0.62. Thus, the best-fit improves over the null hypothesis by $\Delta\,\chi^2 = 4.63$, which for 3 degrees of freedom, corresponds to $p = 0.20$. This is just $0.8\,\sigma$ preference for an oscillatory model over the null hypothesis and is therefore not statistically significant.

In order to derive upper limits on $K$ as a function of ALP mass, we profile the likelihood over the nuisance parameter $\delta_\text{em}$ by maximising the likelihood at each fixed $m_a$ over the parameter $K$ and compare it to the maximal likelihood at the same fixed value of $m_a$. This profile likelihood ratio \cite{Cowan2011, Cowan2013} is computed as,
\begin{equation}
    \lambda_{m_a}(K) = \frac{\mathcal{L}(K, \hat{\hat{\delta}}_\text{em} | m_a)}{\mathcal{L}(\hat{K}, \hat{\delta}_\text{em} | m_a)}
\end{equation}
where $\mathcal{L}(\hat{K}, \hat{\delta}_\text{em} | m_a)$ maximises the likelihood at fixed $m_a$ and $(\hat{K}, \hat{\delta}_\text{em})$ are the corresponding MLEs,
whereas the likelihood $\mathcal{L}(K, \hat{\hat{\delta}}_\text{em} | m_a)$ is evaluated at the MLE of the phase $\hat{\hat{\delta}}_\text{em}$ for $K$ and $m_a$ fixed.
The profile likelihood ratio can be assumed to be $\chi^2$-distributed with a single degree of freedom \cite{Wilks1938}, and can thus provide an upper limit for the parameter $K$. We constructed the test statistics as,
\begin{equation}
    -2 \ln \lambda_{m_a}(K) = \chi^2(K) - \chi^2_\text{min}.
\end{equation}
For one degree of freedom, $-2 \ln \lambda =2.71$ provides the 95\% confidence upper limit. 
Here we apply Wilks theorem \cite{Wilks1938} to compute the upper limit of $K$. As long as $K + \sigma_K \ll 180$\,degree,
the distribution of $K$ can be approximated to be normal. Here, $\sigma_K$ is the dispersion of $K$. To obtain conservative limits on $K$, we imposed the above condition to be satisfied for $K < 60$\,degrees.
When $K > 60$\,degree, we should take the effect of circular statistics into account, see eq.~(\ref{eq:Kcondition}), and a modified von Mises distribution would need to be considered in the likelihood~(\ref{Eq:LikelihoodFn}). 

\begin{figure}
    \centering
    \includegraphics[width=\linewidth]{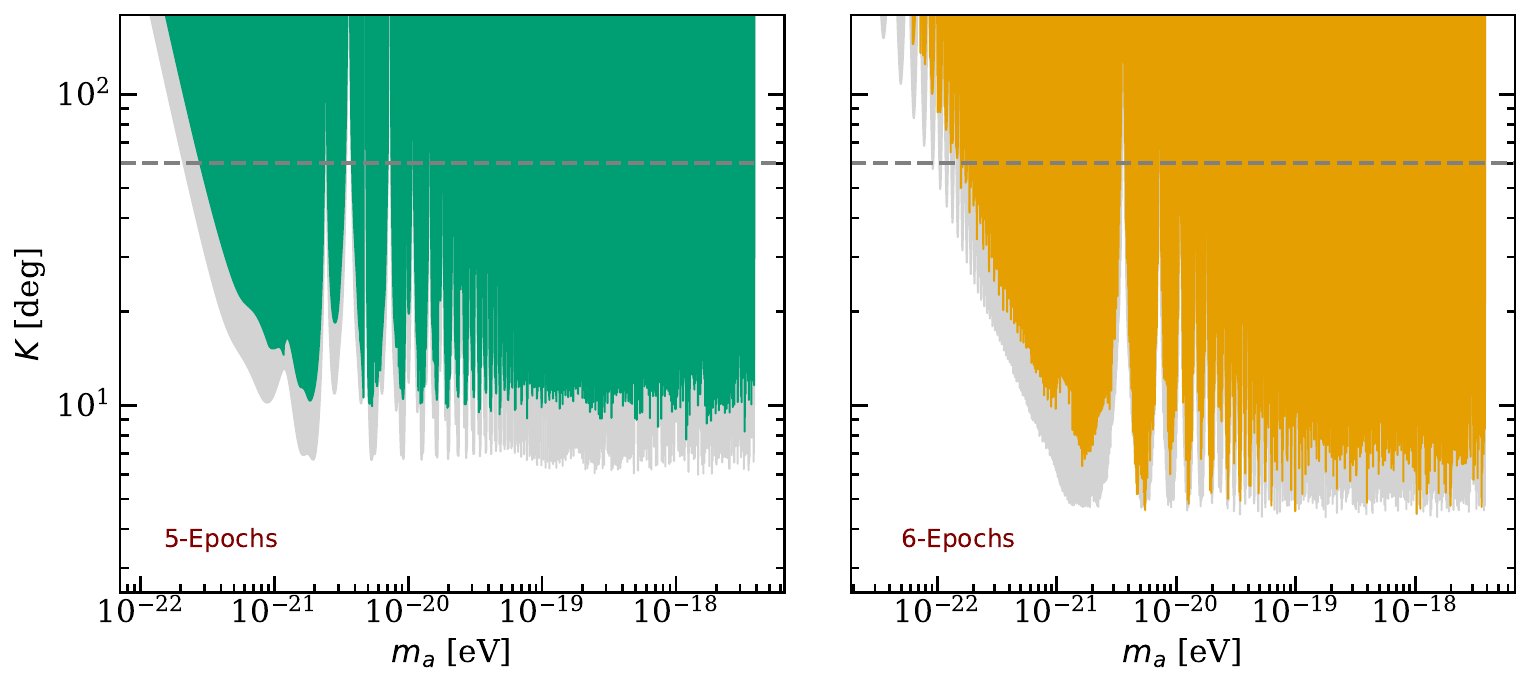}
    \caption{The 95\% exclusion region for parameter $K$, that represents the amplitude of $\Delta\,\theta_{a,{\rm lens}}$, as a function of $m_a$ obtained using profile likelihood. The green region on the \textit{left panel} is computed from combined 5-Epochs, and the orange region on the \textit{right panel} is computed for the 6-Epochs. 
    The grey dashed line at 60\,deg marks the level above which the likelihood ratio may not be approximated by a $\chi^2$-distribution. The grey region shows the potentially accessible region for the given observational set up.}
    \label{fig:K-bound}
\end{figure}

The resulting constraint of $K$ as a function of $m_a$ is shown as a 95\% confidence level (C.L.) upper limit in the left-hand panel of figure~\ref{fig:K-bound} for 5-Epochs, and in the right-hand panel of figure~\ref{fig:K-bound} for 6-Epochs. For the profile likelihood ratio method, there is no need to put any prior information on how low a mass can be probed, 
as it automatically determines the drop of sensitivity at low masses, unlike those discussed in section~\ref{sec:Obs}. 
In order to understand if a stronger limit could have been obtained with the given observational set-up, we also repeat the above procedure setting all measurements to zero while keeping the same measurement uncertainties. This leads to the grey region in figure~\ref{fig:K-bound}. That 
the coloured region is less restrictive for both the 5-Epochs and the 6-Epochs 
is a consequence of the best-fit model providing a better fit than the null hypothesis, however not at a statistically significant level.
The grey-dashed lines in figure~\ref{fig:K-bound} correspond to $K=60$\,degree where the distribution of $K$ is insufficiently described by a normal distribution mentioned above. 
Therefore, when an upper limit on $K$ is later used for constraining $\gag$, we have excluded all $m_a$ that correspond to an upper limit on $K \ge 60$\,degree.

\subsection{Weighted average approach: extension for a sample of lens systems} \label{sec:WtAvg}

Here we compute the weighted differential birefringence angle by combining the measurements of $\Delta\,\theta_{a,{\rm lens}}^{(n)}$.
In order to do this, for a random $\delta_{\rm em}^{(n)}$ in the argument of the sine-term in eq.~(\ref{Eq:DiffBir_multi_incoherent}), at each epoch the sine-term is replaced by it's rms, and $\Delta\,\theta_{a,{\rm lens}}^{(n)}$ for each epoch is given by,
\begin{equation} \label{Eq:DiffBir_wtAvg}
    \Delta \theta_{a,\text{lens}}^{(n)} = \frac{K}{\sqrt{2}} \sin \left(\frac{m_a \Delta t_\mathrm{d}}{2} \right).
\end{equation}
This approximation essentially implies that, at each epoch, the phase of the oscillating ALP field strength are agnostic of each other and thereby $\Delta \theta_{a,\text{lens}}^{(n)}$ are independent of each other.
Under this scenario, we combined $\dthetalens$ by applying directional statistics \cite{Fisher1993, Mardia2000} to compute the weighted average $\langle \dthetalensS\rangle$. 
While this is strictly not applicable for a single lens system, $\langle \dthetalensS\rangle$ provides insight into the efficacy of statistically combining single-epoch data from a sample of lens systems to search for ALP induced birefringence, where $\delta_{\rm em}^{(n)}$ are indeed random. We emphasise that such an analysis with our data only serves as a forecasting exercise. A detailed description for computing $\langle\dthetalensS\rangle$ is presented in appendix~\ref{app:wtAvg}.

The probability distribution function (PDF) of $\langle \dthetalensS\rangle$ for 5- and 6-Epochs are shown as the green and orange shaded histograms in figure~\ref{fig:DiffBir_dist}.
We obtain a 5-Epochs combined $\langle \dthetalensS\rangle = 2.73^{+2.83}_{-2.65}$\,degree (95\% confidence interval), corresponding to 95\% upper limit $|\langle \dthetalensS\rangle| < 5.39$\,degree for $m_a$ in the range $1.2 \times 10^{-21}$\,eV to $3.8\times 10^{-18}$\,eV.
For 6-Epochs, an observed span of almost 9.5\,yr, the range of $m_a$ can be extended towards significantly lighter ALPs with 
$m_a = 2.8 \times 10^{-23}$\,eV,
over which the combined 95\% confidence interval is found to be $\langle \dthetalensS\rangle = 2.21^{+2.95}_{-1.46}$\,degree with a 95\% upper limit $|\langle \dthetalensS \rangle| < 5.16$\,degree.

Note that, while the 
95\% confidence interval on $\langle \dthetalensS\rangle$ is inconsistent with zero, the PDF for 5- and 6-Epochs in figure~\ref{fig:DiffBir_dist} suggests it to be statistically insignificant. As seen in figure~\ref{fig:DiffBir_sinFit}, $\langle \dthetalensS\rangle$ is heavily skewed by $\Delta\,\theta_{a,{\rm lens}}^{(1)}$ measured for $\Delta\,t_{\rm em}^{(1)} = 0$, i.e., by the data on 08 March 2022. 
On discarding $\Delta\,\theta_{a,{\rm lens}}^{(1)}$, the combined $\langle \dthetalensS\rangle$ for the other five epochs is consistent with zero, such that, $\langle \dthetalensS\rangle = 0.42^{+3.67}_{-2.90}$ (95\% confidence interval), and with a 95\% upper limit $|\langle \dthetalensS\rangle| < 4.10$\,degree.

\section{Constraints on ALP from CLASS B1152+199}
\label{sec:constrain_ALP}

\begin{figure}
    \centering
    \includegraphics[width=\linewidth]{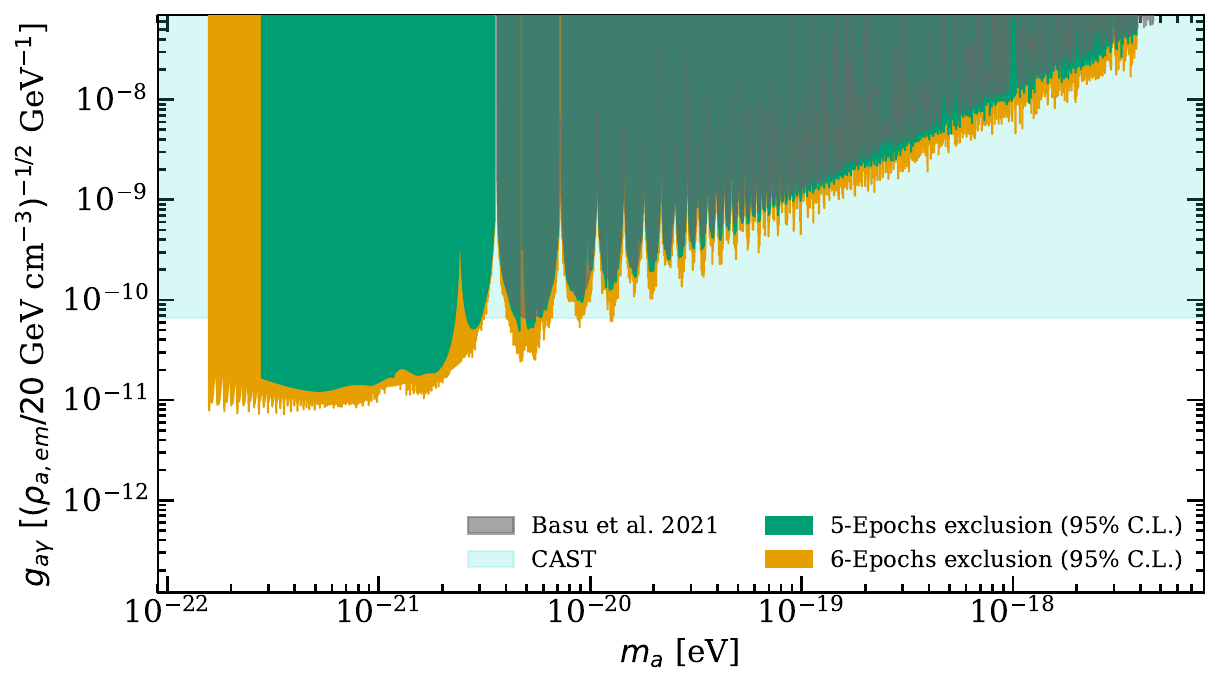}    
    \caption{Constraints on ALP-photon coupling ($\gag$) as a function of ALP mass ($m_a$). The exclusion regions obtained from the 5- and 6-Epochs analysis of B1152+199 are shown as the green and orange regions. The grey shaded region is the exclusion region derived in \cite{Basu2021} at a single epoch. The light blue region is the exclusion region of the CAST collaboration \cite{CAST2017}.
    }
    \label{fig:ExclusionReg}
\end{figure}

In this section, we use the upper limits derived for 
$K(m_a)$ (figure~\ref{fig:K-bound}) in the previous section to constrain the ALP parameter space, $\gag$ as a function of $m_a$. In order to convert the amplitude $K$ to $\gag$, information on the density of the ALP field ($\rho_{a,{\rm em}}$) at the lensed quasar is needed, see eq.~(\ref{Eq:K}). 
For the purpose of this work, we consider ALPs to constitute a part of the dark matter and assume $\rho_{a,{\rm em}}$ to be 50\% of the dark matter density within the Einstein radius of elliptical galaxies that host quasars, and adopt $\rho_{a,{\rm em}} = 20\,{\rm GeV\,cm^{-3}}$ (see \cite{Basu2021} for details). Thus, the values of $\gag$ reported here can be scaled as $\gag\,(\rho_{a,{\rm em}}/{\rm 20\,GeV\,cm^{-3}})^{-1/2}$ for a different value of $\rho_{a,{\rm em}}$.

For deriving limits on $K(m_a)$, different scenarios for the time-coherence of $\delta_{\rm em}^{(n)}$ were considered above. Here we would like to emphasize that, beside time-coherence, spatial coherence of $\delta_{\rm em}^{(n)}$ and $\rho_{a,{\rm em}}$, over which the polarised emission originates in B1152+199, should also be considered \citep[see][]{Basu2021}. The spatial coherence is determined by the de Broglie wavelength ($\lambda_{\rm dB} = 2 \pi/m_a v_a$, with $v_a$ denoting a typical ALP velocity dispersion). For a typical dark matter velocity dispersion of $\sim 100\,{\rm km\,s^{-1}}$ in galaxies hosting quasars, and $m_a \sim 10^{-20}$\,eV, $\lambda_{\rm dB} \approx 30$\,pc. This is similar to the size of the total intensity emitting region of the lensed images as determined through milliarcsec-resolution observations of B1152+199 \cite{Rusin2002}. Thus, we assume $\delta_{\rm em}^{(n)}$ and $\rho_{a,{\rm em}}$ to be spatially coherent in this work.

By considering the scenario that the ALP field strength is coherent over the entire 9.5\,yr duration of the six epochs of observations (eq.~\ref{eq:tau_coherence}), and $\Delta\,\theta_{a,\rm{lens}}^{(n)}$ are related, we 
convert the bound on $K$ in figure~\ref{fig:K-bound} to obtain 
upper limits on $\gag$ using eq.~(\ref{Eq:DiffBir_multi}). 
The resulting exclusion region is shown in figure\,\ref{fig:ExclusionReg}, where the green and the orange shaded regions are the 95\% exclusion obtained for 5- and 6-Epochs, respectively.

The profile likelihood ratio analysis for 5-Epochs allows us to constrain the 
ALP-photon coupling in the mass range,
\begin{equation}
    2.8\times 10^{-22}\;\mathrm{eV}  \leq  m_a \leq 3.8\times 10^{-18}\;\mathrm{eV}, 
\end{equation}
shown as the green exclusion region in figure~\ref{fig:ExclusionReg}.  
The strongest constraint of 
$g_{a\gamma} < 1.2 \times 10^{-11}\;\mathrm{GeV}^{-1} (95\%\ \mathrm{C.L.})$ is found at $m_a = 5.3 \times 10^{-22}\;\mathrm{eV}$.

For the 6-Epochs analysis, 
shown by the orange exclusion region in figure~\ref{fig:ExclusionReg}, we find constraints in the mass range,
\begin{equation}
    1.6 \times 10^{-22}\;\mathrm{eV} \leq m_a \leq  3.8\times 10^{-18}\;\mathrm{eV}.
\end{equation}
The strongest limit is $g_{a\gamma} < 7.2 \times 10^{-12}\;\mathrm{GeV}^{-1} (95\%\ \mathrm{C.L.})$ at $m_a = 3.6 \times 10^{-22}\;\mathrm{eV}$. 
We would like to stress that for some masses the upper limit on the ALP-photon coupling is stronger by more than factor of 10 compared to the analysis in \cite{Basu2021}.

We also determine the upper limit obtained from combined weighted average $|\langle \dthetalensS \rangle|$ to obtain constraints of $\gag$ as a function of $m_a$ using eq.~(\ref{Eq:DiffBir_wtAvg}). The 95\% C.L.\ on the excluded $\gag$ is shown in figure~\ref{fig:ExPlot_wtAvg} in appendix~\ref{app:wtAvg}. As emphasised in section~\ref{sec:WtAvg}, although this method does not consider the coherence of the ALP field oscillation over the observations span, the constraint on $\gag$ improves over the single epoch data \cite{Basu2021} by a factor of 10 wherein the parameter extends towards lower $m_a = 2.8\times 10^{-23}$\,eV. This is mainly because, at lower masses when $2\,\pi/m_a \gg \Delta\,t_{\rm d}$, the term 
$\sin (m_a \Delta\,t_{\rm d}/2)$ in eq.~(\ref{eq:difftheta}) is approximated as $m_a \,\Delta t_{\rm d}/2$. This highlights the fact that, statistical combination of $\dthetalensS$ measured from several lens systems would play an important role in further expanding the ALP-search parameter space.

\section{Discussion and outlook} \label{sec:discussion}

In this paper, we have performed multi-epoch polarimetric observations of the strong gravitational lens system B1152+199 with the VLA (see section~\ref{sec:Obs}) to derive new constraints on ALPs in the ultralight mass regime. For this, we investigated the phenomenon of birefringence induced on linearly polarised synchrotron emission from the lensed quasar due to ALP--photon interaction, to measure the differential birefringence angle $\dthetalens$ between its two lensed images. Using five new observations and an archival data, an observed span of almost 9.5\,yr, allowed us to probe $m_a$ over five orders of magnitude and investigate time-dependent oscillation of ALP field.

\sloppy
In section~\ref{sec:ProfileLikelihood}, we considered that $\dthetalens$ are related via a sinusoidal function across the six epochs for coherence timescale of the ALP field $\tau_{\rm c} \gg \Delta\,t_{\rm mon}$. 
As shown in figure \ref{fig:ExclusionReg}, we 
constrain $\gag \le 7.8 \times 10^{-12}\,(\rho_{a,{\rm em}}/20\,{\rm GeV\,cm^{-3}})^{-1/2}\, {\rm GeV}^{-1}$ to $\le 3.2 \times 10^{-8}\,(\rho_{a,{\rm em}}/20\,{\rm GeV\,cm^{-3}})^{-1/2}\, {\rm GeV}^{-1}$
at 95\% confidence, for $1.6 \times 10^{-22}\,{\rm eV} \le m_a \le 3.8 \times 10^{-18}\,{\rm eV}$. Our measurements improve over the current 95\% confident upper limit of $\gag \le 6.6\times 10^{-11}\,{\rm GeV}^{-1}$ extrapolated to the mass-range of this work obtained with the CERN Axion Solar Telescope (CAST) \cite{CAST2017}. 
Based on WiggleZ galaxy redshift data, the authors of \cite{Hlozek2015} argue that it is plausible that up to 100\% of dark matter density could be attributed to ALPs with $m_a \gtrsim 10^{-24}$\,eV. On the other hand, the comparison of small-scale 1d power spectra from Lyman-$\alpha$ forest observations with corresponding simulations suggest that $m_a > 37.5 \times 10^{-22}$\,eV (at $2\sigma$~C.L.) \cite{Irsic2017}, if ALPS make up 100\% of the dark matter. Competitive constraints on $g_{a\gamma}$ in this mass range probed by gravitational lenses will be crucial to independently rule out or corroborate other constraints. 

For observations of a single lens system and after accounting for the chromatic effects of Faraday rotation, the
only remaining relevant astrophysical parameter is the value of $\Delta\,t_{\rm d}$ 
inferred from modelling the lens system and a measurement of the Hubble rate ($H_0$). 
In the constraint reported above, 
we have assumed them to be known.
Currently, $H_0$ is known at no better than 5\% level, and considering additional uncertainties in the lens model itself, 
a realistic uncertainty on $\Delta\,t_{\rm d}$ is expected to be
5 to 10\%.
A 10\% change in the value of $\Delta\,t_{\rm d}$ leads to a systematic shift by 10\% in the excluded range of $m_a$ and by 5\% on the constrained $g_{a\gamma}$.
Making use of multiple lens systems will allow us to break the degeneracy between $m_a$ and $\Delta\,t_{\rm d}$, apart from the common uncertainty in the Hubble rate. In an upcoming study, we will combine several lens systems, which will then allow us to also marginalise over the uncertainties in $\Delta\,t_{\rm d}$.

Although our reported constraints are consistent with the null hypothesis, 
the possibility of a mild oscillatory behaviour of $\dthetalens$ from our data cannot be ruled out as seen in figure~\ref{fig:DiffBir_sinFit}. 
The variation of differential birefringence angle with time, when modelled as a sinusoidal
oscillation, see eq.~(\ref{Eq:DiffBir_multi}), produces a fit passing through all the measurement points,
corresponding to $m_a \approx 1.5\times 10^{-21}$\,eV and $\gag \approx 1.3\times 10^{-11}\,(\rho_{a,{\rm em}}/20\,{\rm GeV\,cm^{-3}})^{-1/2} \;\mathrm{GeV}^{-1}$. However, the significance of this fit over the null hypothesis remains below $1\,\sigma$. 
The method of combined analysis of multi-epoch measurements of $\dthetalens$ demonstrates how it could allow not only to obtain upper limits on $\gag$, but also measure almost all relevant parameters of an ALP and extend to a sample of lens systems. For multi-epoch data obtained from a sample of lens systems, $m_a$ and $\gag$ should agree on the same value, leaving us with measurements of the ratios of ALP density between different lens systems.
Only a single overall ALP density would remain as an unknown, which would need 
complementary measurements.

Using a simple approach of weighted average of $\dthetalens$ 
described in section~\ref{sec:WtAvg}, we constrain $|\langle \dthetalens\rangle| \le 5.16$\,degree at 95\% confidence. As emphasised earlier, weighted average is only a pedagogical method to gauge the efficacy of combining $\dthetalensS$ measured for a sample of lens systems. The constraint obtained from our data, corresponds to a 95\% upper-limit of $\gag \le 8.3\times10^{-12} \,(\rho_{a,{\rm em}}/20\,{\rm GeV\,cm^{-3}})^{-1/2}\, {\rm GeV}^{-1}$ to $\le 2.8\times 10^{-8}\,(\rho_{a,{\rm em}}/20\, {\rm GeV\, cm^{-3}})^{-1/2}\, {\rm GeV}^{-1}$ for $2.8\times 10^{-23}\,{\rm eV} \le m_a \le 3.8\times 10^{-18}\,{\rm eV}$ (figure~\ref{fig:ExPlot_wtAvg}). While simple averaging ignores the coherence of ALP field strength and its phase, 
nonetheless, this provides insight into improving constraints on $\dthetalens$ when measurements from multiple lens systems are combined. 
Combination of six data points improved over the single epoch measurement $\Delta\,\theta_{a,{\rm lens}}^{(6)}$ \cite{Basu2021} by about 40\%. However, when $\Delta\,\theta_{a,{\rm lens}}^{(1)}$ is disregarded in our analysis (see section~\ref{sec:WtAvg}), an improvement of $\approx 1/\sqrt{5}$ is seen for the remaining five epochs. This suggests that statistically combining $\dthetalensS$ from a sample of lens system can probe significantly lower $\gag$. We have refrained from quoting constraint on $\gag$ without $\Delta\,\theta_{a,{\rm lens}}^{(1)}$.

As the method of using differential birefringence and its combination from multi-epoch and/or multiple lens systems to search for ALPs predominantly depends on the level of statistical error, there is a realistic chance of detecting such a signal if ALPs exist. An improvement in the search parameter space using six measurements of $\dthetalensS$ is already seen over a single epoch measurement with low number statistics presented in this work. However, the expected statistical improvement is mostly limited by a single measurement as pointed out above. Note that,
assuming a Gaussian distribution, $\Delta\,\theta_{a,{\rm lens}}^{(1)}$ 
agrees with the null hypothesis (no achromatic birefingence) with a probability of $p = 0.02$. Using the binomial distribution to estimate the probability of finding one such measurement out of six,
turns out to be $\approx 10\%$. While we cannot exclude that the oscillating behaviour of $\dthetalens$ is real (figure~\ref{fig:DiffBir_sinFit}), 
epoch 1 may indeed be a statistical outlier. It is also possible that there are underlying systematics at low-level. Hence, longer follow up monitoring of several lens systems are necessary to confidently rule out anomalies with low number statistics.

Our constraints are competitive with other 
observations and experiments
that investigate alternative signatures of ALPs in the ultralight mass regime. 
Our study improves over the CAST constraint on $\gag$ by up to an order of magnitude for $1.6\times 10^{-22}\,{\rm eV} \le m_a \le 3\times 10^{-21}\,{\rm eV}$.
Other significant constraints obtained through astrophysical measurements in this mass regime is provided through constraining X-ray spectral distortions caused by photon--ALP interconversion in the presence of magnetic fields of galaxy clusters that host luminous AGNs \cite{Berg2017, Reynolds2020, Sisk-Reynes2022}. Observations with the \textit{Chandra} X-ray Observatory of such a signal constrained $\gag < 6.3\times 10^{-13}\,{\rm GeV}^{-1}$ for $m_a < 10^{-12}$\,eV at 99.7\% C.L \cite{Sisk-Reynes2022}. However, the nature of the signal, and therefore the derived limits, heavily depend on the assumed strength and morphology of the magnetic fields, and free-electron density along the line of sight. In contrast, our limits only make an assumption on $\rho_{a,{\rm em}}$, which however, does not affect the properties of the birefringence signal itself. Recently, birefringence signal was searched through polarimetric measurements of pulsars in the European Pulsar Timing Array, that constrained $\gag < 2\times 10^{-13}\,{\rm GeV}^{-1}$ to $\gag \lesssim
10^{-11}\,{\rm GeV}^{-1}$ at 95\% C.L. in the $m_a$ range $3\times 10^{-23}$\,eV to $5\times10^{-21}$\,eV for $\rho_{a,{\rm em}} \approx 0.4\,{\rm GeV\,cm^{-3}}$ \citep{Porayko2025}. 
Measuring birefringence signal from pulsars are susceptible to systematics arising from Faraday rotation in the ionosphere and varying positional angle of the polarised emission from pulsars \cite{Porayko2025}. Whereas, differential birefringence measurements presented here are largely unaffected by such effects. Currently, constraints on cosmic birefringence from observations of polarised CMB supersede all other limits for the mass range considered in this work with $g_{a\gamma} < 9.6 \times 10^{-13}\, \mathrm{GeV}^{-1}$ for $m_a = 10^{-12}\, \mathrm{eV}$ and $g_{a\gamma} \propto m_a$ \cite{Fedderke2019}. 
However, CMB measurements are limited by cosmic variance, a natural limit beyond which CMB measurements are insensitive to the parameter space. In contrast, with sensitive telescopes and observations of multiple lens systems at multiple epochs, we are not aware of a principle sensitivity limit on ALP parameters for the method presented in \cite{Basu2021} and further refined in this work.

The previous constraint obtained on ALPs using B1152+199 \citep{Basu2021}, the method of $QU$-fitting \citep{Mao2017}, a parametric modeling of the Stokes\,$Q$ and $U$ spectra, was used to derotate the observed polarisation angles of the lensed images A and B, and estimate $\Delta\,\theta_{a,{\rm lens}}^{(6)}$. In contrast, here we employ the non-parametric technique of RM synthesis, that provides $\dthetalens$ on a similar level of accuracy (see figure~\ref{fig:DiffBir_sinFit}). For individual lens systems, measurement accuracy of $\dthetalensS$ is predominantly limited by the level of statistical uncertainties, especially in the measurement of RM used for derotating the observed angle of linear polarisation (see eq.~\ref{eq:theta0}). Improvement in measurements of RM can be achieved through a combination of sensitive observations \citep{iacobelli2013}, and/or broader frequency coverage, especially towards low frequencies \citep{Brentjens&Bruyn2005, Schnitzeler2015}. On one hand, in terms of sensitivity, the advantage of RM synthesis over $QU$-fitting is that, RM synthesis provides polarisation parameters at $\nu_{\rm eff}$ that has the coherent sensitivity of the entire observed bandwidth. While, in order to robustly perform $QU$-fitting, a sensitivity $\gtrsim2\,\sigma$ at each channel is required. This often leads to significantly longer telescope time. Hence, application of RM synthesis to polarised strong gravitational lenses detected in large sky-area polarimetric surveys, e.g., surveys with the upcoming SKA-Mid in the frequency range 4.6 to 10\,GHz (Band\,5 receivers) \citep{McKean2015} would allow us to search for ALPs effectively. Furthermore, RM synthesis opens up the possibility of detecting fainter polarised lens systems compared to B1152+199 in such surveys and significantly improve constraints on $\dthetalensS$ by statistically combing them (see section~\ref{sec:WtAvg}). On the other hand, for individual lens systems, better measurement of RM and distinction of imprints of complex polarisation spectrum on Faraday depth spectrum (see appendix~\ref{app:error}) can be made at low frequencies \citep[e.g.,][]{Schnitzeler2015, VanEck2017, VanEck2018, basu2019, OSullivan2023LoTSS}, e.g., $\lesssim 0.5$\,GHz, which is otherwise difficult to achieve on individual systems with observations at gigahertz frequencies even through deep observations. In this regard, observations of strongly lensed systems with the upgraded LOFAR2.0, with its international stations, can provide high resolution polarimetric data near 0.15\,GHz \citep{Hut2024LOFAR2} and significantly improve measurements of $\dthetalensS$. In summary, future telescopes like the SKA-Mid and LOFAR2.0 can play a significant role in expanding the ALP search parameter space.

In this work we have ignored the so-called `washout effect' that has been considered when analysing birefringence signal from polarised 
CMB \cite{Fedderke2019}, and in astrophysical systems \cite{Chen2022a, Chen2022b}.
This effect is caused due to different amount of ALP-induced birefringence from oscillating ALP field when the polarised emission originate at different time because of finite light travel time within the emitting volume. Ignoring the washout effect means that we implicitly make the assumption that the emission region is not only coherently oscillating in time, but also compact enough to guarantee spatial coherence. This is a plausible assumption, but it must be tested eventually. In order to appropriately account for any potential washout effect, a detailed model for the emission region and its properties along the line of sight 
is required. This can be addressed through high-angular resolution observations on milli-arcsec scales using very large baseline interferometry, and will be addressed in a future study.

\section{Conclusion} \label{sec:conclusion}

We have established multi-epoch measurements of differential birefringence angle ($\dthetalensS$) from multiply, unresolved lensed images in strong gravitational lens systems to be a powerful probe of ultralight ALP field at emission. For this, broad-bandwidth, polarimetric observations of the strongly lensed system B1152+199, lensed into two unresolved images, was used. Using a non-parametric approach of applying RM synthesis for determining RM to mitigate the effect of chromatic Faraday rotation, we obtain a similar level of accuracy in measuring $\dthetalensS$ as that from parametric fitting of the Stokes\,$Q$ and $U$ spectra. This implies, RM synthesis can be efficaciously applied to large sky-area polarimetric surveys to detect faint polarised lensed systems and use them to search for ALPs. Using data from six epochs spanning over 9.5\,yr, we improve the 95\% confidence exclusion region of $\gag$ by more than 10 times, and extend the excluded ALP mass $m_a$ towards lower mass by almost two order of magnitude with respect to that of a single epoch measurement of $\dthetalensS$ of B1152+199. Our study also significantly improves over the bound on $\gag$ obtained with the CAST by up to an order of magnitude.

Multi-epoch measurements of $\dthetalensS$ have the potential to discover ALPs, provided they exist. If a discovery is made for a lens system, this method has the advantage of further consistency check through a precise prediction of the signal in another lens system. Future large sky-area surveys with the SKA-Mid and LOFAR2.0 can significantly expand the ALP search parameter-space through statistical combination of $\dthetalensS$ obtained from several lens systems. For individual lens systems, deeper and longer monitoring campaigns, especially at low frequencies, will also boost ALP searches and alleviate anomalies with low number statistics and/or any underlying systematics at low level.

\acknowledgments

The National Radio Astronomy Observatory (NRAO) and Green Bank Observatory are facilities of the U.S. National Science Foundation operated under cooperative agreement by Associated Universities, Inc. We acknowledge financial support from BMFTR-Verbundforschung ErUM-Pro grants D-MeerKAT-II and D-MeerKAT-III (grant numbers 05A20PBA and 05A23PBA) and from NRW-MKW Profilbildung 2020 project Big Bang to Big Data (B3D). A.~B. gratefully acknowledges the financial support provided by the German Federal Ministry of Research, Technology and Space (BMFTR) in the framework of the Knowledge creates perspectives for the region!, for the project StStG -- DZA -- Aufbauphase: Deutsches Zentrum für Astrophysik, Großforschungszentrum in der sächsischen Lausitz: Aufbauphase 2026, grant number 03WSP1746. Y.~U. is supported by Grant-in-Aid for Scientific Research under Contract Nos. JP21KK0050, JP23K25873 (23H01177), JP26H00402(26H00402), and JST FOREST Program under Contract No. JPMJFR222Y. S.~D. acknowledges the VLA Data Reduction Workshop 2022 for the warm hospitality and in-depth insights into radio interferometry. We thank the support provided by the NRAO CASA helpdesk for helping us to address the continuity of bandpass across VLA's L-, S- and C-band receivers. 
We thank E.\,Ros, S.\,Schumacher, T.\,Kovacs, M.\,Oguri, K.\,Ichiki, N.\,Kitajima, J.\,Goswami and M.\,Sato for their help during the preparation of the observing proposal with the VLA. This research has made use of
\texttt{Astropy},\footnote{\url{http://www.astropy.org}} a community-developed core Python
package for Astronomy \citep{astropy:2013, astropy:2018}, \texttt{NumPy}
\citep{numpy11}, \texttt{LMFIT} \citep{lmfit2025} and \texttt{Matplotlib} \citep{matplotlib07}.

\paragraph{Data availability.} The radio continuum visibility data obtained with the VLA for this work are available through the NRAO Data Archive (\url{https://data.nrao.edu/portal/}) under the project code 22A-059 (PI: A.\,Basu).

\section*{Appendix}
\appendix

\section{Stokes $Q$ and $U$ spectra of CLASS B1152+199} \label{app:QUspec}

\begin{figure}
    \centering
    \begin{tabular}{cc}
    \includegraphics[width=0.5\linewidth]{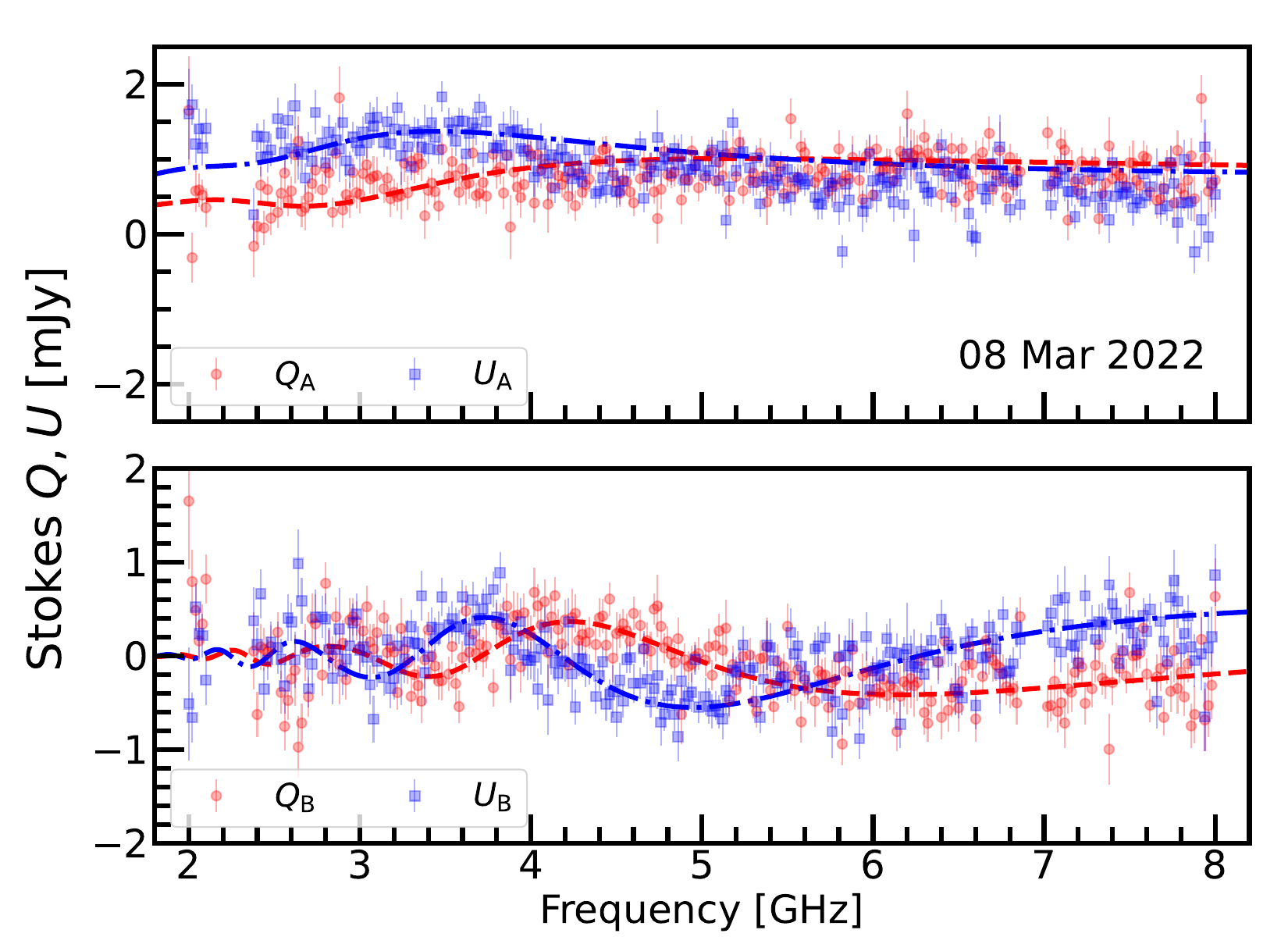}&
    \includegraphics[width=0.5\linewidth]{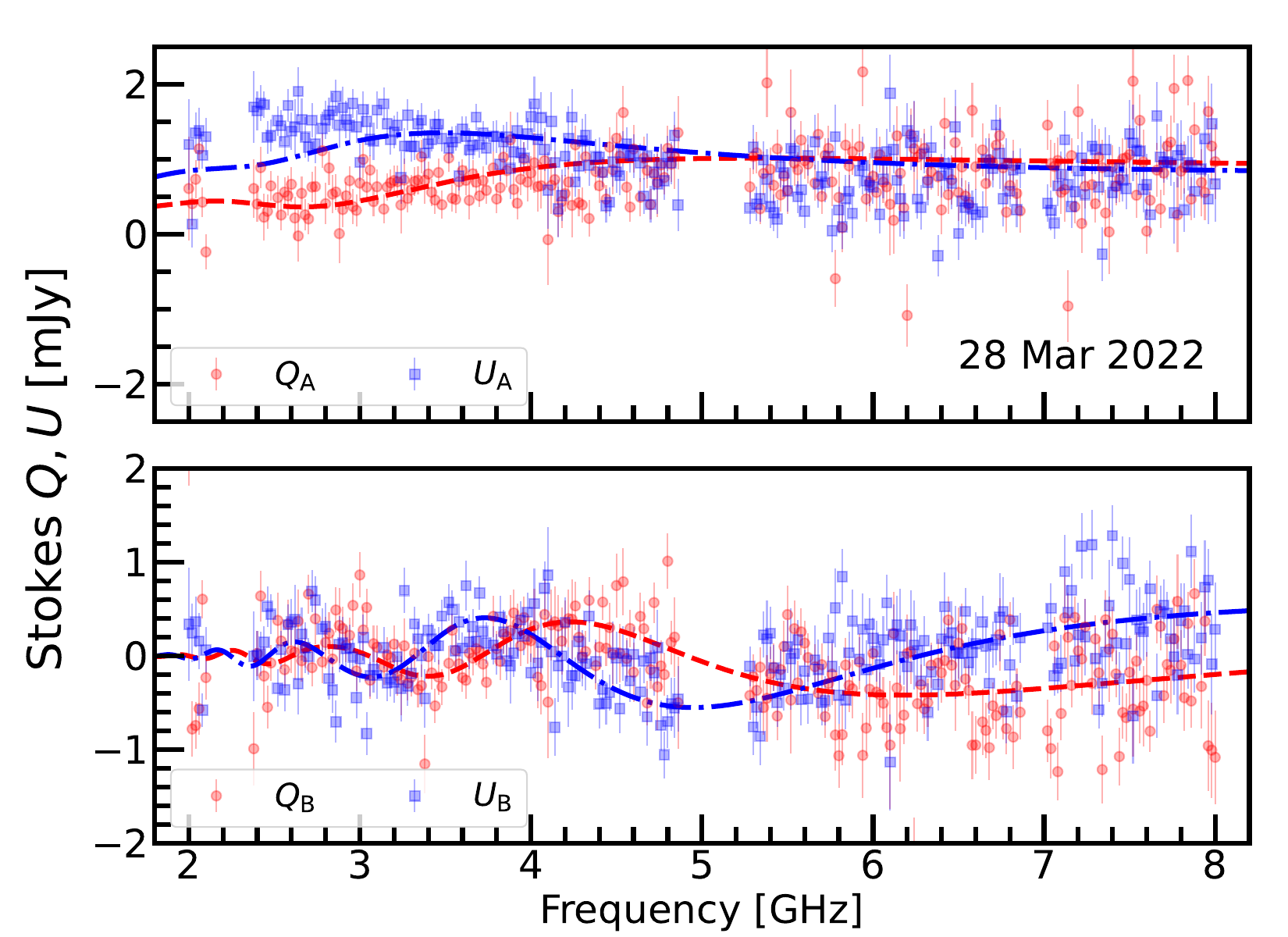}\\
    \includegraphics[width=0.5\linewidth]{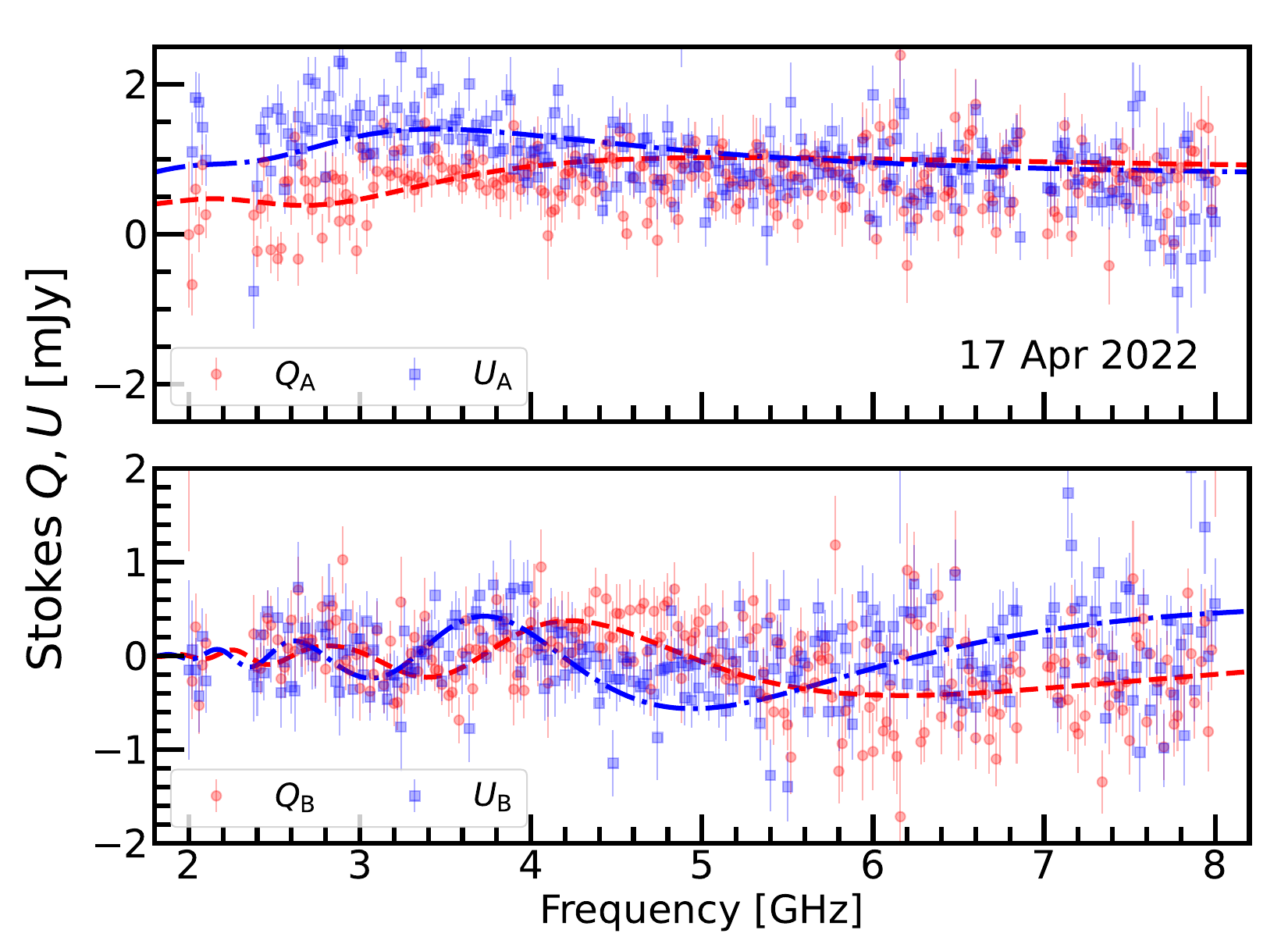}&
    \includegraphics[width=0.5\linewidth]{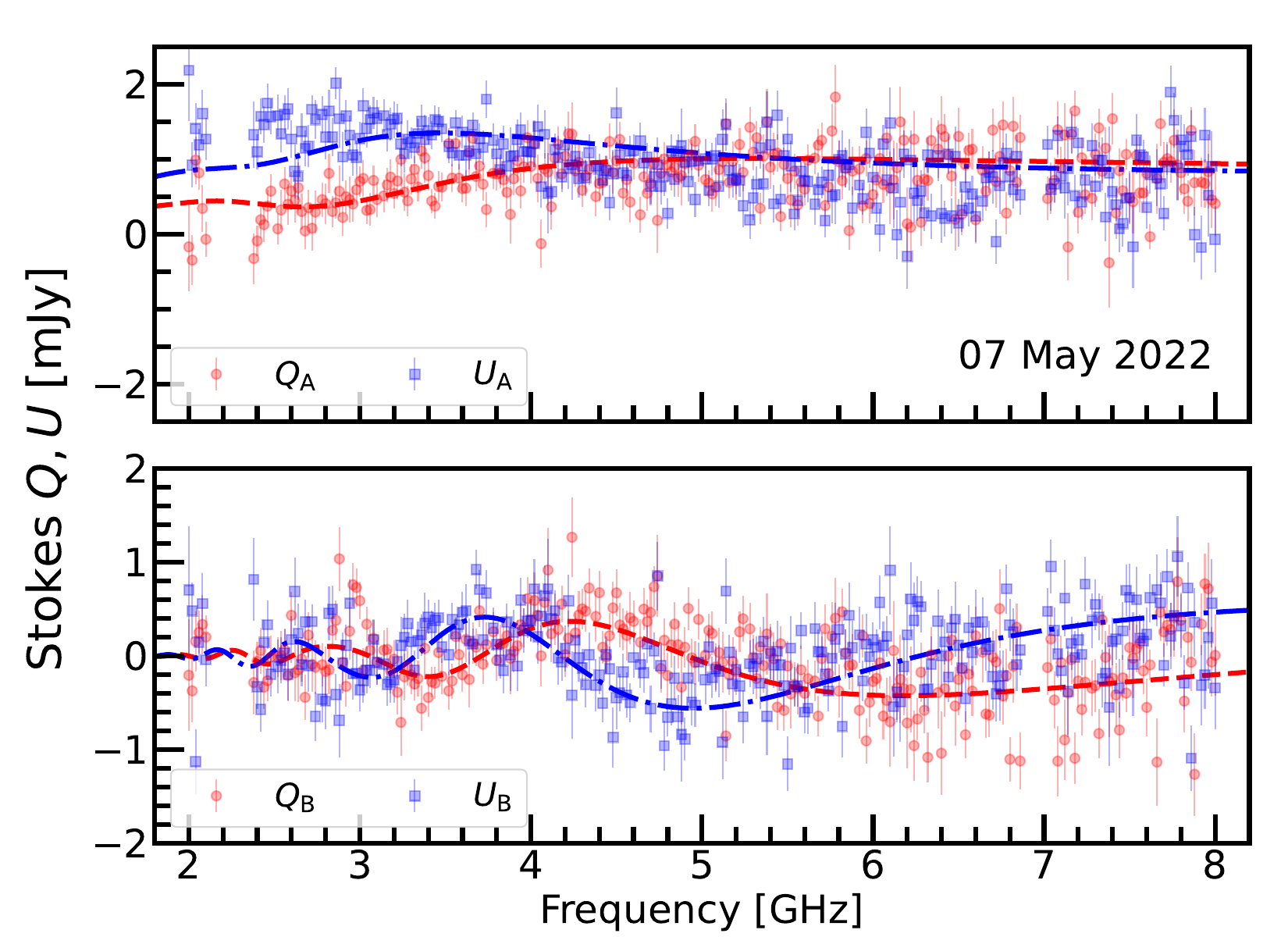}\\
    \includegraphics[width=0.5\linewidth]{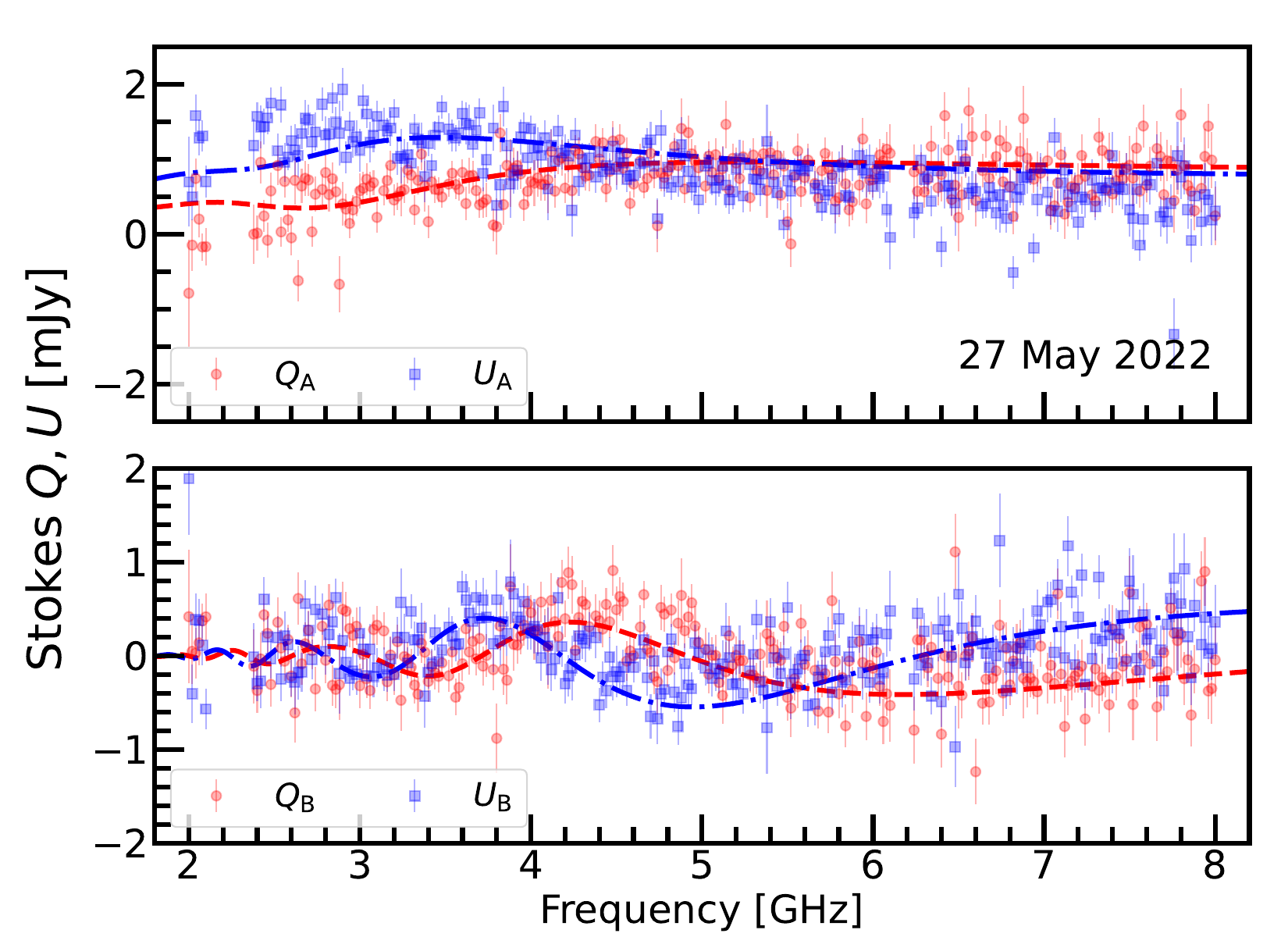}&
    \\
    \end{tabular}
    \caption{Spectrum of the Stokes\,$Q$ (red circles) and $U$ (blue squares) parameters obtained for 5-Epochs. The top and bottom rows in each of the panels are for Image A and B, respectively. The dashed red and dash-dotted blue lines represent the modelled Stokes\,$Q$ and $U$ parameters using $QU$-fitting taken from Mao et al. \cite{Mao2017}.}
    \label{fig:QUspec}
\end{figure}

Here we present the Stokes\,$Q,U$ spectrum obtained from the 5-Epochs data listed in table~\ref{tab:epochs} for the lensed images A and B in B1152+199. As described in section~\ref{sec:PolAna}, the flux densities of Stokes\,$Q$ and $U$ of each of the lensed images at each 20-MHz channel image were obtained by modelling them as unresolved 2-D Gaussian in the plane of the sky. The different panels in figure~\ref{fig:QUspec} show the Stokes\,$Q$ (red circles) and $U$ (blue squares) flux densities as a function of frequency for different epoch of observations. The top and bottom rows in each panel are for image A and B, respectively. The modelled Stokes\,$Q,U$ spectra obtained from the technique of $QU$-fitting \citep{Mao2017} are shown as the red dashed and blue dash-dotted lines. In order to obtain the model in flux density units, we multiplied the model presented for fractional $q = Q/I$ and $u = U/I$ in \cite{Mao2017} by the fitted Stokes\,$I$ spectrum at respective epoch from our data presented in table~\ref{tab:flux}.

\section{Estimation of uncertainties} \label{app:error}

Analyses of linear polarisation parameters, and the method of RM synthesis undergoes non-linear processing that makes error estimation via standard error propagation methods rather challenging. Furthermore, robust estimation of error of RM and Stokes\,$Q(\nu_{\rm eff}), U(\nu_{\rm eff})$ are imperative as they directly impact 
the differential birefringence angle, $\dthetalens$. A standard means of computing an error on RM ($\rm \delta\,RM$) is given by the empirical relation $\delta\,{\rm RM} = {\rm RMSF}/(2\times{\rm SNR})$ 
\cite{iacobelli2013}, where, ${\rm RMSF} \approx 2\sqrt{3}\,[1/\nu_{\rm min}^2 - 1/\nu_{\rm max}^2]/c^2$, $\nu_{\rm min}$ and $\nu_{\rm max}$ are the minimum and maximum frequency of the observations, and SNR is the signal to noise ratio of the polarised intensity. Such a relation to compute $\delta\,{\rm RM}$ is inadequate when the Faraday depth spectrum is insufficient to be described by polarised emission originating at a single RM. The polarised emission from both 
images in B1152+199 have significantly complicated frequency dependence. Modelling of the Stokes\,$Q,U$ spectrum suggests at least two polarised components in each of the lensed images \cite{Mao2017}. This manifests as the broad component and sub-structure \cite{basu2019} predominantly seen in the Faraday depth spectrum of fainter Image\,B, and as slightly broader base for Image\,A in figure~\ref{Fig:RMspectrum}. Therefore, in order to estimate RM and $\delta\,\rm RM$, we re-sample the Stokes\,$Q,U$ spectrum shown in figure~\ref{fig:QUspec} at each epoch.
Here, Stokes\,$Q,U$ flux densities at each 20-MHz frequency channels were randomly drawn within the respective rms noise of each 20-MHz channel image, and RM synthesis was performed. This was repeated 100 times. The mean and standard deviation of these random draws were used to compute Stokes\,$Q(\nu_{\rm eff})$ and $U(\nu_{\rm eff})$ reported in table~\ref{tab:polarization_AB}, and $\rm RM_A$ and $\rm RM_B$ in table~\ref{tab:rm_theta_AB}.

In order to determine the error on $\theta_{\rm 0,A}, \theta_{\rm 0,B}$ and $\dthetalens$, we performed Monte Carlo simulations starting with $5\times10^4$ realisations of Stokes\,$Q(\nu_{\rm eff})$, $U(\nu_{\rm eff})$ and RM for each of the epochs, and images\,A and B separately. 
Each of these quantities were sampled from Gaussian distribution to determine the probability distribution functions (PDF) of $\theta_{\rm 0,A}$ and $\theta_{\rm 0,B}$ using eq.~(\ref{eq:theta0}), and they were subsequently used to determine the PDF of $\dthetalens$ using eq.~(\ref{eq:difftheta}), shown as histograms in figure~\ref{fig:DiffBir_dist}. We note that the PDFs of $\Delta \theta_{a,\text{lens}}$ are well-approximated by normal distribution as shown by the solid lines. We use the 68\% confidence interval around the median $\dthetalens$ for each epoch, listed in table~\ref{tab:rm_theta_AB}, for further analyses.
We emphasise that, because measurement of synchrotron polarisation is sensitive to the orientation and not its direction, $\theta_{\rm 0,A}$ and $\theta_{\rm 0,B}$ are periodic between $-\pi/2$ to $\pi/2$. Therefore, the values of $\dthetalens$ for each epoch lies within $[0,\pi]$ or $[-\pi,0]$ depending on whether $\theta_{\rm 0,A} > \theta_{\rm 0,B}$ or $\theta_{\rm 0,A} < \theta_{\rm 0,B}$, respectively. We would like to point out that, because the error on RM depends on $\nu_{\rm min}$ and $\nu_{\rm max}$ of the data used for RM synthesis, the confidence interval on $\dthetalens$ is between 1--4\,degree worse for the 2--8\,GHz frequency range used by us instead of the entire 1--8\,GHz.

\begin{figure}
    \includegraphics[width=0.8\linewidth]{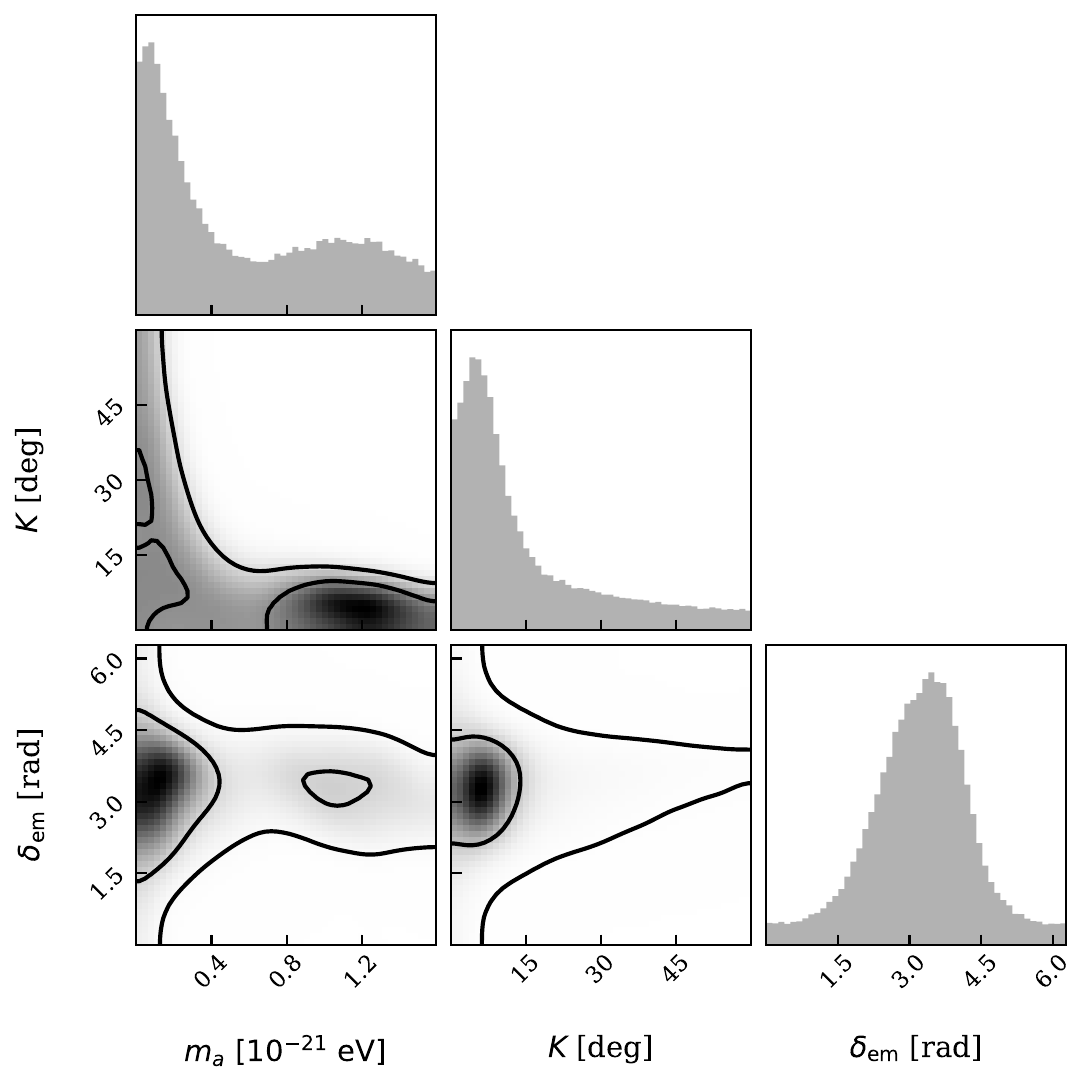}    
    \caption{Posterior distribution of parameters for 6-Epochs computed using Markov Chain Monte Carlo (MCMC) method. The contours represent 1 and 2$\sigma$ levels.}
    \label{fig:CornerPlot}
\end{figure}

\section{Posterior distribution of parameters}\label{app:Posterior}

We perform a Bayesian parameter estimation of the three-parameter model $(m_a, K, \delta_\mathrm{em})$ defined in eq.~(\ref{Eq:DiffBir_multi}) using Markov Chain Monte Carlo (MCMC) sampling implemented with the \texttt{emcee} \cite{emcee} package in python. 
Here, we characterise the posterior geometry of parameter space to confirm that the data are consistent with the null hypothesis of no ALP signal. 
We adopt the Gaussian likelihood given in eq.~(\ref{Eq:LikelihoodFn}), and assign uniform priors to all three parameters. The prior ranges are physically motivated as described in eqs.~(\ref{eq:Kcondition}) -- (\ref{eq:mass}).
However, as detailed in section \ref{sec:ProfileLikelihood}, 
the prior on $K$ is limited to $60$\,degree, and further
limit the upper bound of $m_a$ to $2\pi/(3\,t_{\rm cad}) \approx 2\pi/30 \; {\rm day^{-1}}$ 
to evaluate lowest-frequency, non-trivial solutions.

The sampler is run with 16 walkers for 100,000 steps each. 
The integrated autocorrelation time $\tau_\mathrm{int}$ for each parameter is computed to assess adequate sampling of the posterior. The condition $\tau_\mathrm{int} \ll (\mathrm{chain \; length})/50$ is well satisfied.
We discard the first 500 steps as burn-in and thin the chain by a factor of 15.

The marginalised 1D and 2D posterior distributions for $(m_a, K, \delta_\mathrm{em})$ are shown in figure \ref{fig:CornerPlot}. The data do not prefer a specific non-zero value for parameters $m_a$ and $K$, and both posteriors are consistent with upper limits rather than measurements. Due to epoch 1, the posterior phase shows a mode at $\delta_\mathrm{em} \sim \pi$. However, a genuine detection of an ALP-induced birefringence oscillation would be expected to produce a localised posterior in all three parameters.

\begin{figure}
    \centering
    \includegraphics[width=\linewidth]{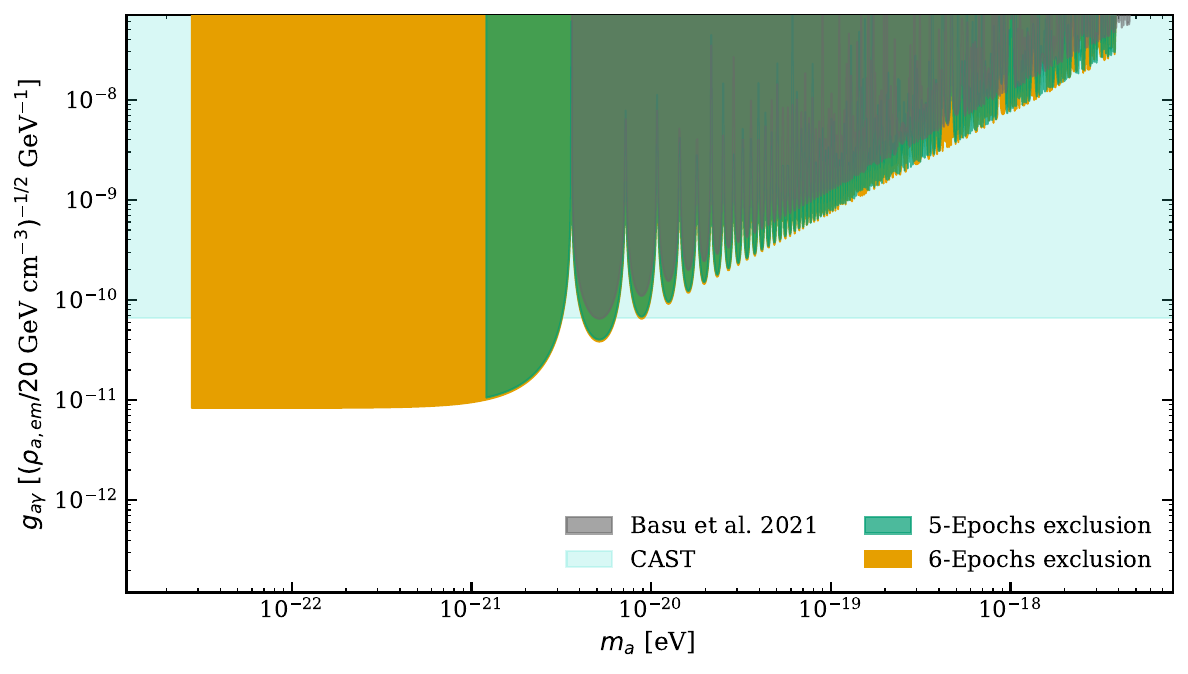}
    \caption{The 95\% excluded region for $\gag$ obtained using five epoch monitoring of B1152+199 is shown as the shaded green region. The orange region shows the bound on $\gag$ for six epoch spanning over 9.5\,yr. These bounds are representative of combination of a sample of lenses. For comparison, we show the bound presented in Basu et al. \cite{Basu2021} using a single epoch observations of B1152+199 as the grey shaded region.}
    \label{fig:ExPlot_wtAvg}
\end{figure}

\section{Constraints on ALPs from weighted average differential birefringence} \label{app:wtAvg}

The combined weighted differential birefringence angle, $\langle\dthetalensS\rangle$, is computed by applying directional statistics \cite{Fisher1993, Mardia2000},
\begin{equation}
    \langle \dthetalensS \rangle = \arctan\left(\frac{\sum\limits^{5}_{n = 1} w_n\, \sin \Delta\,\theta_{a,\rm{lens}}^{(n)}}{\sum\limits^{5}_{n = 1} w_{n}\,\cos \Delta\,\theta_{a,\rm{lens}}^{(n)}} \right)
\end{equation}
Here, $w_n$ are the weights, given by the inverse of 68\% width of the PDF of $\Delta\,\theta_{a,\rm{lens}}^{(n)}$ in figure~\ref{fig:DiffBir_dist}.
As there are only five (six) measurements for 5-Epochs (6-Epochs), we have performed a Monte-Carlo simulation with 50\,000 random samples of $\Delta\,\theta_{a,{\rm lens}}^{(n)}$, each drawn from a modified von Mises distribution that ensures the random angles are sampled within [$-\pi/2$, $+\pi/2$] (see \cite{Basu2021}) with their respective angle given in table~\ref{tab:rm_theta_AB} and corresponding 68\% interval as the concentration parameter. 

In figure~\ref{fig:ExPlot_wtAvg}, we show the 95\% upper limit on $\gag$ as a function of $m_a$ 
using constraints on $|\langle \dthetalensS\rangle|$ reported in section~\ref{sec:WtAvg}.
The green and orange shaded regions are for 5-Epochs and 6-Epochs, respectively. As discussed in section~\ref{sec:WtAvg}, this exclusion plot is strictly not applicable to B1152+199, but provides an overview of how statistical combination of many lensed systems in a similar way can improve the ALP search parameter space. In such a case, the minimum $m_a$ that can be probed is determined by the lens system that has the largest $\Delta\,t_{\rm d}$ between its lensed images \cite{Basu2021}.

\bibliography{references}

\end{document}